\documentclass[12pt]{article}
\usepackage{graphicx,amssymb,amsmath}
\usepackage{epsfig}
\usepackage{color}
\usepackage{cite}
\usepackage{multirow}
\newcommand\fverb{\setbox\pippobox=\hbox\bgroup\verb}
\newcommand\fverbdo{\egroup\medskip\noindent%
                              \fbox{\unhbox\pippobox}\ }
\newcommand\fverbit{\egroup\item[\fbox{\unhbox\pippobox}]}
\newbox\pippobox
\newcommand{\beq} {\begin{equation}}
\newcommand{\eeq} {\end{equation}}
\newcommand{\beqa} {\begin{eqnarray}}
\newcommand{\eeqa} {\end{eqnarray}}

\newcommand{\be}{\begin{equation}}
\newcommand{\ee}{\end{equation}}
\newcommand{\bea}{\begin{eqnarray}}
\newcommand{\eea}{\end{eqnarray}}

\def\om{\omega}

\begin{document}
 
\begin{flushright}
HIP-2017-21/TH \\
NSF-ITP-17-093
\end{flushright}

\begin{center}

\centerline{\Large {\bf Holographic pinning}}

\vspace{8mm}

\renewcommand\thefootnote{\mbox{$\fnsymbol{footnote}$}}
Niko Jokela,${}^{1,2}$\footnote{niko.jokela@helsinki.fi}
Matti J\"arvinen,${}^3$\footnote{jarvinen@lpt.ens.fr} and
Matthew Lippert${}^{4,5}$\footnote{Matthew.Lippert@liu.edu}

\vspace{4mm}
${}^1${\small \sl Department of Physics} and ${}^2${\small \sl Helsinki Institute of Physics} \\
{\small \sl P.O.Box 64} \\
{\small \sl FIN-00014 University of Helsinki, Finland} 

\vspace{2mm}
${}^3${\small \sl Institut de Physique Th\'eorique Philippe Meyer \&\\ Laboratoire de Physique Th\'eorique,\\
\'Ecole Normale Sup\'erieure, PSL Research University, CNRS,\\ 24 rue Lhomond, 75005 Paris, France \\
}

\vspace{2mm}
\vskip 0.2cm
${}^4${\small \sl Department of Physics} \\
{\small \sl Long Island University} \\
{\small \sl Brooklyn, New York, USA} 

\vspace{2mm}
\vskip 0.2cm
${}^5${\small \sl Kavli Institute for Theoretical Physics} \\
{\small \sl University of California} \\
{\small \sl Santa Barbara, California, USA} 

\end{center}

\vspace{8mm}

\setcounter{footnote}{0}
\renewcommand\thefootnote{\mbox{\arabic{footnote}}}

\begin{abstract}
\noindent
In a holographic probe-brane model exhibiting a spontaneously spatially modulated ground state, we introduce explicit sources of symmetry breaking in the form of ionic and antiferromagnetic lattices.
For the first time in a holographic model, we demonstrate pinning, in which the translational Goldstone mode is lifted by the introduction of explicit sources of translational symmetry breaking.  The numerically computed optical conductivity fits very well to a Drude-Lorentz model with a small residual metallicity, precisely matching analytic formulas for the DC conductivity. 
We also find an instability of the striped phase in the presence of a large-amplitude ionic lattice.
\end{abstract}

\newpage
\tableofcontents

%%%%%%%%%%%%%%%%%%%%%%%%%%%%%%%%%%%%%%%%%%%%%%%%%%%%%%%%%%%%%%%%%%%%%%%%%%%%%%%%%%%%%%%%%%%%%%%%%%%%%%
%%%%%%%%%%%%%%%%%%%%%%%%%%%%%%%%%%%%%%%%%%%%%%%%%%%%%%%%%%%%%%%%%%%%%%%%%%%%%%%%%%%%%%%%%%%%%%%%%%%%%%
%%%%%%%%%%%%%%%%%%%%%%%%%%%%%%%%%%%%%%%%%%%%%%%%%%%%%%%%%%%%%%%%%%%%%%%%%%%%%%%%%%%%%%%%%%%%%%%%%%%%%%
%%%%%%%%%%%%%%%%%%%%%%%%%%%%%%%%%%%%%%%%%%%%%%%%%%%%%%%%%%%%%%%%%%%%%%%%%%%%%%%%%%%%%%%%%%%%%%%%%%%%%%

\newpage
\section{Introduction}

Electronic systems at low temperature exhibit a range of phases with spontaneously broken translation symmetries.  A charge density wave (CDW) is a typical example, in which continuous translation symmetry is broken in one direction, leading to stripes of modulated charge density.\footnote{For a review, see \cite{Gruner}.}  Other order parameters can become spatially modulated, such as spin density waves (SDW) or persistent circulating currents.  Complex combinations of intertwined orders can occur, such as the pair-density waves (PDW) found in certain underdoped cuprates, featuring spatially modulated charge, spin, and superconducting phase.

Spontaneously striped phases have interesting properties and rich dynamics.  The charge conductivity, in particular, can feature striking  behavior.  Because the symmetry breaking is spontaneous, striped phases feature a Goldstone mode which is the translation zero-mode.  An applied electric field can easily cause the stripes to slide, resulting in collective charge transport.

In practice, however, additional sources of explicit symmetry breaking, such as impurities or the underlying lattice, typically generate a spatially-varying potential for the stripes and lift the Goldstone mode.  The stripes are then pinned in place and can only slide if the potential barrier is overcome by a sufficiently large electric field.  This depinning transition results in a highly nonlinear conductivity.

Spontaneous striped order is a common feature of many condensed matter systems.  In some cases, the underlying physics can be understood in a weakly interacting, quasiparticle description.  However, other examples are found in strongly coupled materials, such as the pseudogap regime of cuprate superconductors \cite{Tranquada, Fujita}, and are more amenable to a holographic approach.

Holographic modeling of striped phases has been a subject of much attention in recent years.  In most of the examples, translation symmetry is broken explicitly.  A spatially modulated chemical potential can represent an ionic lattice \cite{Horowitz:2012ky,Donos:2013eha}, and a sum of such potentials with different frequencies and random phases can model disorder \cite{Arean:2013mta}.

Comparatively less attention has focused on the more interesting case of spontaneous symmetry breaking.  Striped phases of several holographic models featuring non-perturbative spontaneous striped order have been constructed; see e.g., \cite{Ooguri:2010kt, Ooguri:2010xs, Bayona:2011ab, Donos:2012gg, BallonBayona:2012wx, Bu:2012mq, Rozali:2012es, Donos:2012yu,Donos:2013wia, Withers:2013loa, Withers:2013kva, Rozali:2013ama, Ling:2014saa, Jokela:2014dba, Donos:2016hsd, Amoretti:2016bxs,Cremonini:2017usb}.  In a few cases, phases with spontaneous order in two directions have been found \cite{Withers:2014sja, Donos:2015eew, Cai:2017qdz}. However, the unique transport properties of these states are just beginning to be studied. 

In \cite{Jokela:2016xuy}, we initiated the analysis of the conductivity of a spontaneously striped state.  We focused on the D3-D7' probe-brane model, a well-studied holographic model featuring a spontaneously striped phase with modulated magnetization, persistent transverse currents, and modulated charge density.  Phases with such complex intertwined orders appear in, for example, cuprate \cite{Berg} and ferrous superconductors \cite{Balatsky}, and have recently also been modeled holographically, for example, in \cite{Cremonini:2017usb}. We found that the stripes slide with a velocity proportional to the applied electric field and carry a significant fraction of the current.

In experimental systems, spontaneous breaking coexists with disorder or an underlying lattice.  Several recent studies have investigated how explicit symmetry breaking affects the formation of holographic striped phases.  In \cite{Andrade:2015iyf, Cremonini:2017usb}, background linear scalars were shown to impact the modulated instability of the homogeneous phase and wavelength of the resulting stripes.  Even more interestingly, \cite{Andrade:2017leb} demonstrated that an explicit ionic lattice of sufficient amplitude can force the wavelength of the modulated instability to be half-integer multiples of the lattice wavelength, indicating a commensurate lock-in between the lattice and the stripes.  However, the effects of an explicit lattice on the conductivity, especially the pinning of holographic stripes, the focus of this paper, have not previously been investigated.\footnote{Pinning in a holographic CDW was reported in \cite{Ling:2014saa}.  However, as the model lacks an explicit source of symmetry breaking to lift the Goldstone mode, we disagree with the identification of the reported conductivity as being due to pinning.}

In this paper, we analyze the linear conductivity of the D3-D7' model with the addition of explicit translation symmetry breaking, either in the form of an modulated chemical potential (ionic lattice) or a background antiferromagnetic field (magnetic lattice).  This explicit breaking lifts the Goldstone mode and pins the stripes.  The resulting longitudinal conductivity is well fit by a Drude-Lorentz model.  We find a small residual DC conductivity, computed semi-analytically in terms of background horizon data, which represents the current of charge carriers flowing across both the stripes and the lattice.

The transverse conductivity is relatively unaffected by the addition of the explicit lattice and is still well fit by a Drude form.  As a result, the surprising approximate symmetry between the longitudinal and transverse DC conductivities found in \cite{Jokela:2016xuy} is strongly broken.  The DC conductivity across the stripes is now an order of magnitude smaller than along the stripes.

The DC Hall conductivity in the absence of a lattice \cite{Jokela:2016xuy} features a delta peak due to the  persistent transverse current oscillating as stripes slide.  Adding a lattice pins the stripes and regulates this delta peak into a modified Lorentzian form, with the same resonance frequency and relaxation time as seen in the longitudinal conductivity.

In many respects the ionic and magnetic lattices have qualitatively similar effects.  However, because the charge modulation of the stripes is subleading, the potential well due to the ionic lattice is shallower.  As a result, the resonant frequency is an order of magnitude smaller than for the magnetic case.

However, one surprising result is that an ionic lattice of sufficient amplitude leads to an instability.  We find that, as the lattice amplitude is increased, at a certain point the resonant frequency goes to zero and then becomes imaginary.  This is a sign that a pole in the current-current correlator has acquired a positive imaginary part, corresponding to an exponentially growing pseudo-Goldstone mode.

The rest of this paper is organized as follows.  In Sec.~\ref{sec:set-up}, we will review the construction of the D3-D7' model and the spontaneously spatially modulated phase.  Sec.~\ref{sec:introducing_lattices} introduces the explicit modulation.  Then, in Sec.~\ref{sec:conductivities}, we recompute the conductivities, first for the magnetic lattice in Sec.~\ref{sec:magnetic_lattice} and then ionic lattice in Sec.~\ref{sec:ionic_lattice}.  We conclude with a summary and open questions in Sec.~\ref{sec:summary}.

\emph{Note added}: Two papers on closely related topics are appearing concurrently with this one. Ref.~\cite{Andrade:2017cnc} investigates the conductivity and pinning of a spontaneously striped phase of a holographic Bianchi VII construction in the presence of an explicit lattice.  Ref.~\cite{Alberte:2017cch} studies transverse, gapped pseudo-phonons in the context of a holographic massive gravity model.

%%%%%%%%%%%%%%%%%%%%%%%%%%%%%%%%%%%%%%%%%%%%%%%%%%%%%%%%%%%%%%%%%%%%%%%%%%%%%%%%%%%%%%%%%%%%%%%%%%%%%%%%%%%%%%%%%%%%%%%%%%%%%%
%%%%%%%%%%%%%%%%%%%%%%%%%%%%%%%%%%%%%%%%%%%%%%%%%%%%%%%%%%%%%%%%%%%%%%%%%%%%%%%%%%%%%%%%%%%%%%%%%%%%%%%%%%%%%%%%%%%%%%%%%%%%%%
%%%%%%%%%%%%%%%%%%%%%%%%%%%%%%%%%%%%%%%%%%%%%%%%%%%%%%%%%%%%%%%%%%%%%%%%%%%%%%%%%%%%%%%%%%%%%%%%%%%%%%%%%%%%%%%%%%%%%%%%%%%%%%
%%%%%%%%%%%%%%%%%%%%%%%%%%%%%%%%%%%%%%%%%%%%%%%%%%%%%%%%%%%%%%%%%%%%%%%%%%%%%%%%%%%%%%%%%%%%%%%%%%%%%%%%%%%%%%%%%%%%%%%%%%%%%%
%%%%%%%%%%%%%%%%%%%%%%%%%%%%%%%%%%%%%%%%%%%%%%%%%%%%%%%%%%%%%%%%%%%%%%%%%%%%%%%%%%%%%%%%%%%%%%%%%%%%%%%%%%%%%%%%%%%%%%%%%%%%%%

\section{Set-up}
\label{sec:set-up}
The D3-D7' model is a holographic model of strongly interacting fermions on a $(2+1)$-dimensional defect \cite{Bergman:2010gm}, which in many ways resembles graphene \cite{Davis:2011gi, Omid:2012vy, Semenoff:2012xu}.  The construction consists of a probe D7-brane embedded in a D3-brane background such that supersymmetry is completely broken and stabilized by internal magnetic fluxes wrapping 2-cycles in the $S^5$.  We will only briefly review the relevant aspects of the model here; for more details, see \cite{Jokela:2016xuy}.  It is noteworthy to mention other closely related holographic constructions \cite{KeskiVakkuri:2008eb,Alanen:2009cn,Jokela:2011eb,Jokela:2011sw,Kim:2011zd,Jokela:2012se,Lippert:2014jma,Bea:2014yda,Mezzalira:2015vzn}, which have significantly contributed to the understanding of the current setting.

The probe D7-brane is embedded so that it spans the $t,x,$ and $y$ Minkowski directions, is extended in the holographic radial direction $r$, and wraps both of the 2-spheres. The bulk solutions are specified by the embedding functions $z$ and $\psi$ and the worldvolume gauge field $a_\mu$. The D7-brane action consists of a Dirac-Born-Infeld term and a Chern-Simons term:
\be 
 S  =  -T_7 \int d^8x\, e^{-\Phi} \sqrt{-\mbox{det}(g_{\mu\nu}+ 2\pi\alpha' F_{\mu\nu})} -\frac{(2\pi\alpha')^2T_7}{2} \int P[C_4]\wedge F \wedge F \ . \label{totalact}
\ee
From this action, we obtain equations of motion for the embedding fields $\psi$ and $z$ as well as the worldvolume gauge fields $a_t$, $a_x$, and $a_y$.\footnote{The equation for the radial component $a_u$ gives a constraint enforcing the radial gauge condition $a_u = 0$.}

We scale out the temperature $T$ by rescaling the spatial coordinates $x^\mu$ and gauge field $a_\mu$ by the horizon radius $r_T$.  We furthermore work with a compact radial coordinate 
\be
 u = \frac{r_T}{r} \ ,
\ee
which sets the location of the horizon at $u=1$ and the anti-de Sitter (AdS) boundary at $u=0$.

We can include a chemical potential $\mu$ and magnetic field $b$ by turning on appropriate components of the bulk gauge field $a_\mu$.  For a particular ratio of charge density to magnetic field, the D7-brane can take a Minkowski embedding, which is holographically dual to a gapped, quantum Hall phase \cite{Bergman:2010gm,Jokela:2010nu,Jokela:2013hta,Jokela:2014wsa}.  However, we will concentrate in this paper on the generic black hole embedding, which is dual to a gapless quantum fluid. We will further restrict to embeddings with zero fermion mass.

\subsection{Spontaneous stripes}\label{sec:spontaneous_stripes}

At large chemical potential and small magnetic field, the homogeneous solution of the D3-D7' model was found to exhibit an instability at nonzero momentum \cite{Bergman:2011rf, Jokela:2012vn}.  In \cite{Jokela:2014dba}, this instability was shown to lead to a striped ground state, featuring spatially modulated charge density, magnetization, and persistent current along the stripes.  Rotation invariance allows us to choose the modulation to be in the $x$ direction, while translation symmetry is preserved in the $y$ direction.

The spontaneous modulation has a dynamically determined spatial frequency $k_0$, which is an increasing function of $\mu$ and a decreasing function of $b$.  The transverse gauge field $a_y$ and the embedding $\psi$ exhibit the leading modulation. Because all the bulk fields are nontrivially coupled, these induce a subleading modulation with a frequency $2k_0$ in the temporal gauge field $a_t$ and embedding scalar $z$. The spatial period of the solution is $L = 2\pi/k_0$.

In \cite{Jokela:2016xuy}, we analyzed the electrical conductivity for this striped state at zero magnetic field.  We computed the DC conductivity $\sigma^\mathrm{DC}$ semi-analytically in terms of horizon data using the procedure of \cite{Donos:2015gia, Banks:2015wha, Donos:2015bxe, Araujo:2015hna}, and the AC conductivity $\sigma(\omega)$ was computed numerically.

Our most significant result was that the stripes move as a result of an electric field $E_x$ applied across the stripes.  The stripes have a sliding mode which is the Goldstone mode of the spontaneously broken translation symmetry. In the absence of an underlying lattice or localized impurities explicitly breaking translation invariance, the stripes are not pinned to any particular location.  The nonbackreacted D3-brane sector acts as a momentum sink, analogous to uniformly smeared impurities, providing friction to the sliding stripes.  This results in a sliding velocity $v_s$ proportional to $E_x$ and a finite $\sigma_{xx}^\mathrm{DC}$.

The optical conductivity in both the $x$ and $y$ directions exhibits a Drude form at low frequency. Surprisingly, the conductivity across the stripes and along the stripes is equal to within a few percent, despite the anisotropy of the background and very different mechanisms for charge transport in the two directions.

The Hall conductivity $\sigma_{yx}$ illustrates the effect of the sliding stripes.  The striped ground state features a modulated transverse persistent current $j_y(x)$, which slides along with the stripes.  The transverse current at a fixed location therefore oscillates in time as the stripes slide past.  This results in a delta peak in the DC Hall conductivity with a modulated strength.

\subsection{Introducing lattices}\label{sec:introducing_lattices}

In this paper, we add explicit translation symmetry breaking on top of the nonlinear spontaneous striping.  As the striped background consists of both modulated magnetization and charge density, there are two interesting ways we can introduce explicit translation symmetry breaking.  We can consider modulation either in the background magnetic field or in the chemical potential.  We refer to the resulting field theory configurations as magnetic and ionic lattices, respectively.  And, note that we only introduce one type of lattice at a time, not both simultaneously. 

In a sufficiently strong periodic potential, striped states will typically adjust their wavelength to be commensurate with that of the potential.  Such commensurate lock-in has been observed in a holographic context in \cite{Andrade:2017leb}, and we expect such an effect to occur in this model.  However, we leave this investigation for future work.  

In the meantime, we set the wavelength of the lattice equal to the wavelength dynamically preferred by the stripes in the absence of a lattice. The spatial frequency of the magnetic lattice is then $k_0$ to match the spontaneous magnetization of the stripes, and the ionic lattice is given a frequency $2k_0$ to match the stripes' charge density modulation.  We furthermore choose the phase of the lattice such that the discrete symmetries of the striped solution are respected.  Essentially, we impose commensurability of the lattice and the stripes by hand.

These lattices are dual in the bulk to modulated boundary conditions for the dual components of the world volume gauge field.  For the two lattices, the boundary conditions that we introduce are then as follows:
\bea
\label{magnetic_boundary_condition}
{\rm{Magnetic\ lattice\ :\ }} \qquad a_y(x,u=0) & = & b x + \alpha_b \sin(k_0 x) \\
\label{ionic_boundary_condition}
{\rm{Ionic\ lattice\ :\ }} \qquad a_t(x,u=0) & = & \mu + \alpha_\mu \cos(2k_0 x) \ .
\eea
The parameters $\alpha_b$ and $\alpha_\mu$ measure the amplitude of the lattices and will play a major role in the subsequent analysis.\footnote{We restrict to positive $\alpha$. Notice that the sign of both $\alpha$'s can be changed by choosing the phase of the modulation differently, so formally the boundary conditions with opposite signs are equivalent. Separate branches of solutions obtained by a deformation of the $\alpha=0$ solution towards negative $\alpha$ do exist, however, but only for $|\alpha| \ll 1$. We expect that they are subdominant even in the narrow range where they exist.}

In this paper, we will restrict our attention to solutions for which the spatially averaged background magnetic field vanishes, $b = 0$.  We further set the chemical potential to be $\mu=4$, which is sufficiently large that the system is in the spontaneously striped phase \cite{Jokela:2014dba}.\footnote{In fact, we only know definitively  that the system is in the striped phase for $\mu=4$ when $\alpha_b = \alpha_\mu= 0$.  We postpone a detailed analysis of the phase diagram in the presence of explicit lattices to future work.}

\begin{figure}[!ht]
\center
 \includegraphics[width=0.50\textwidth]{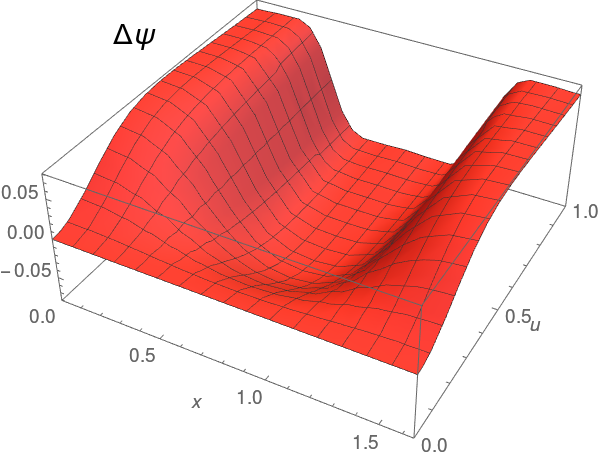}%
 \includegraphics[width=0.50\textwidth]{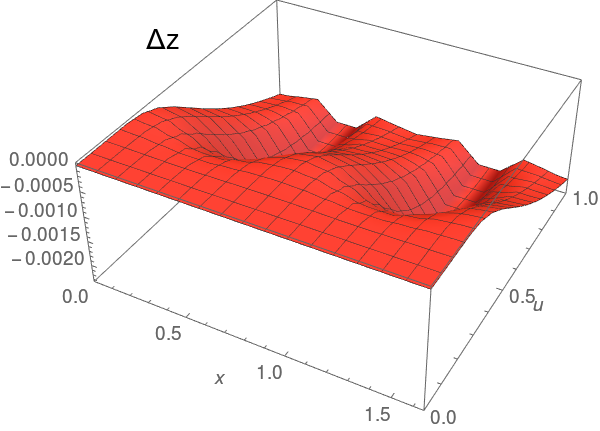}
 \includegraphics[width=0.50\textwidth]{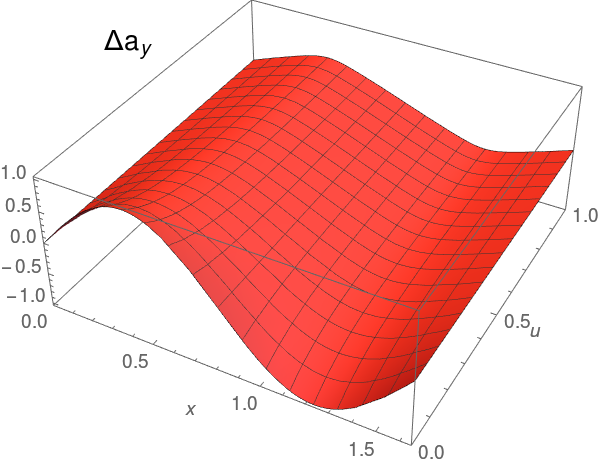}%
 \includegraphics[width=0.50\textwidth]{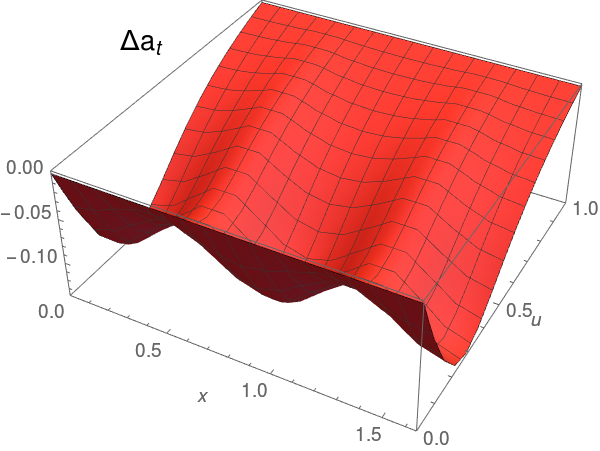}
 \caption{The difference between the bulk stripe solution with a magnetic lattice $\alpha_b = 1$ solution and without $\alpha_b = 0$.  Top left: $\Delta \psi$, top right: $\Delta z$, bottom left: $\Delta a_y$, and bottom right: $\Delta a_t$.} 
  \label{fig:magnetic_lattice_solution}
\end{figure}

Before moving on to study the conductivity, we note that the lattices have a significant effect on the striped background itself and not just on its excitations.  We show in Fig.~\ref{fig:magnetic_lattice_solution} the difference between the striped solution with a magnetic lattice $\alpha_b=1$ and without $\alpha_b=0$.  
Not surprisingly, the magnetic lattice directly and strongly enhances the modulation of $a_y$. The various couplings between all the fields induce smaller enhanced modulations in $\psi$, $z$, and $a_t$.  
Just as the spontaneous modulation of $z$ and $a_t$ is twice the frequency of $a_y$ and $\psi$, so too is the enhanced modulation generated by the lattice.
Also note that, although $\Delta a_t$ is substantial, one should recall that, since the boundary condition sets $a_t(u=0) = \mu = 4$, the fractional effect $\Delta a_t/a_t$ is actually smaller than for $\psi$.

\begin{figure}[!ht]
\center
 \includegraphics[width=0.50\textwidth]{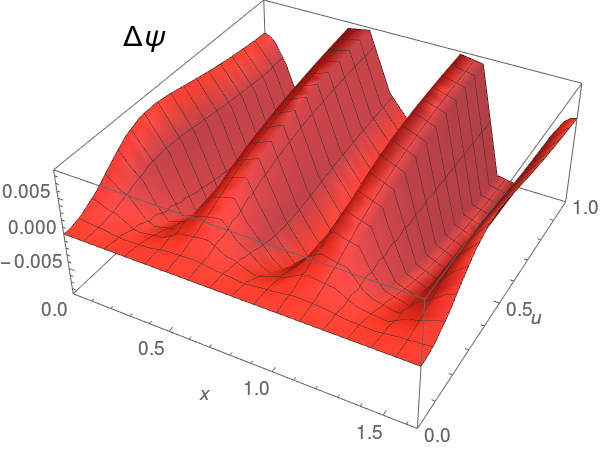}%
 \includegraphics[width=0.50\textwidth]{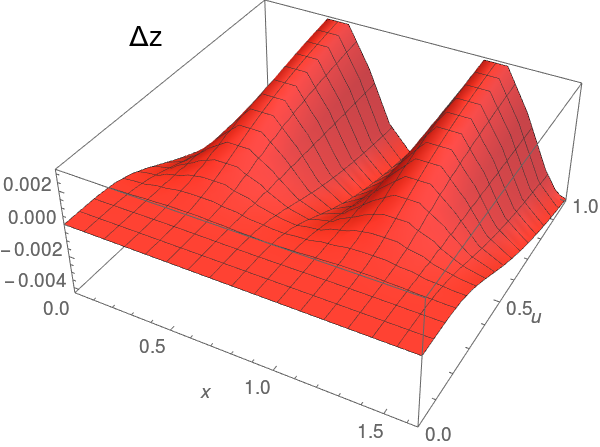}
 \includegraphics[width=0.50\textwidth]{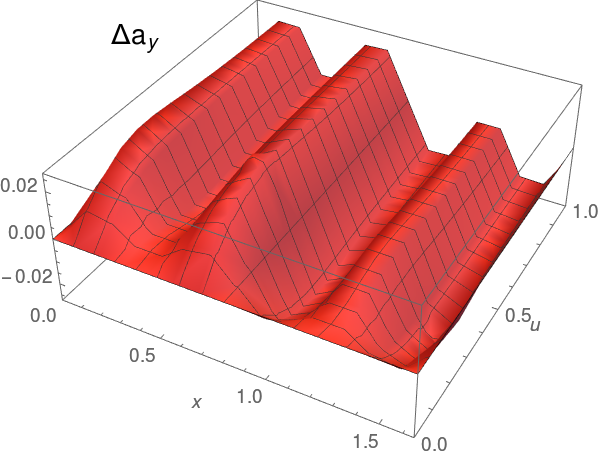}%
 \includegraphics[width=0.50\textwidth]{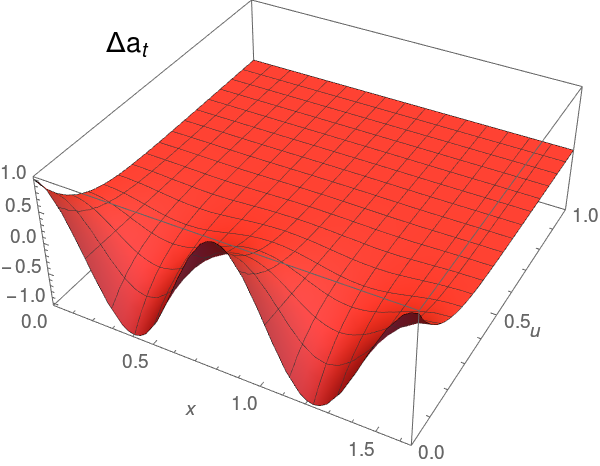}
 \caption{The difference between $\alpha_\mu = 1$ solution and $\alpha_\mu = 0$. Top left: $\Delta \psi$, top right: $\Delta z$, bottom left: $\Delta a_y$, and bottom right: $\Delta a_t$.} 
  \label{fig:ionic_lattice_solution}
\end{figure}

In Fig.~\ref{fig:ionic_lattice_solution}, we show the analogous plot for an ionic lattice, the difference between the striped solutions with $\alpha_\mu = 1$ and $\alpha_\mu = 0$.  Here, the lattice directly generates a sizable increase in the modulation of $a_t$, which is then transmitted to the other fields. 
Even though we have in both cases chosen the smallest wave number for the source which preserves all discrete symmetries, there are also clear differences in the shape of the response between Fig.~\ref{fig:magnetic_lattice_solution} and Fig.~\ref{fig:ionic_lattice_solution}. 
In particular, for $\psi$ and $a_y$, the ionic lattice primarily enhances the amplitude of the mode with frequency $3k_0$ rather than $k_0$.

\begin{figure}[!ht]
\center
 \includegraphics[width=0.50\textwidth]{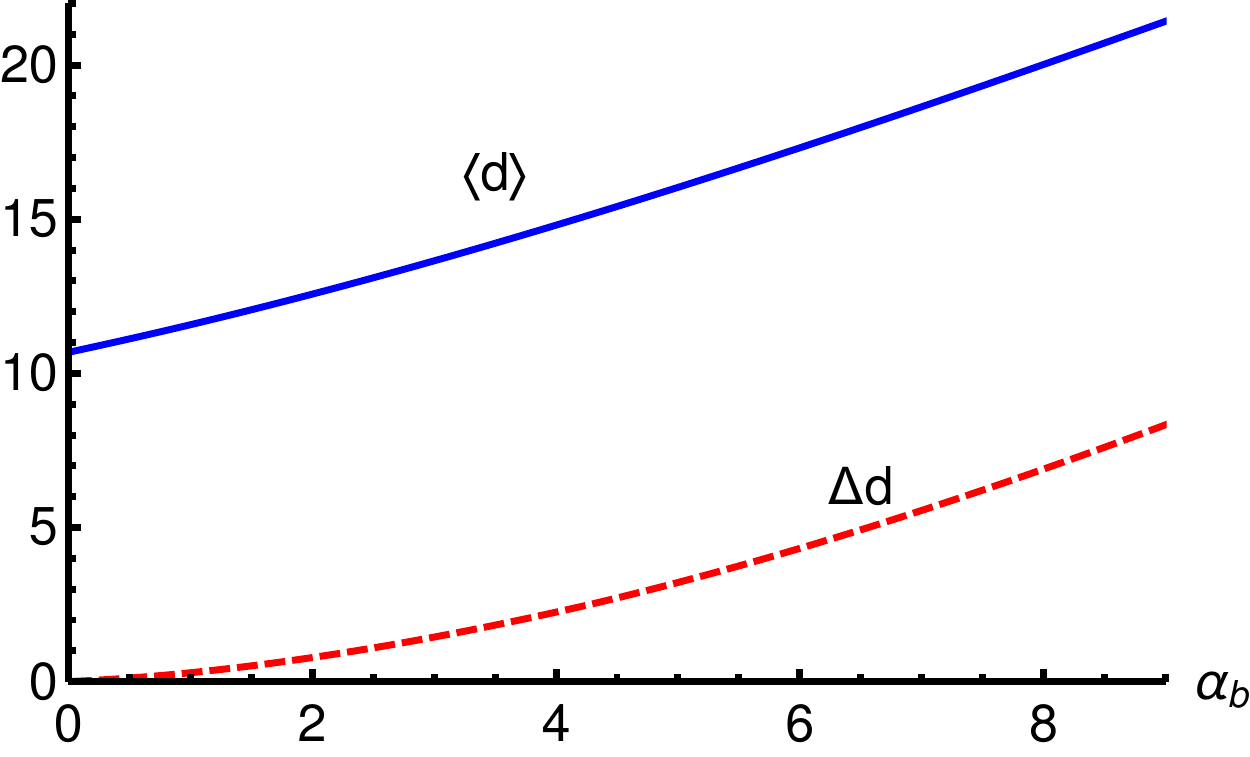}% 
 \includegraphics[width=0.50\textwidth]{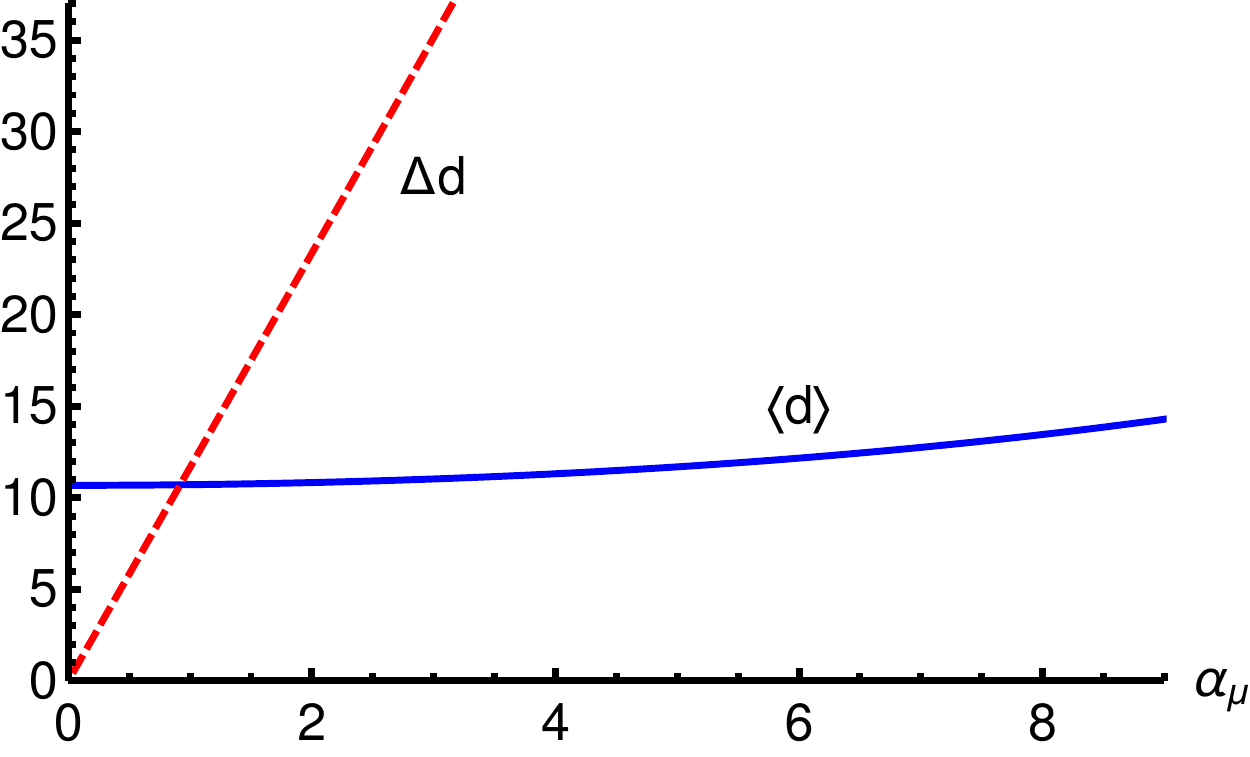}
 \caption{The spatial average of the charge density $\langle d\rangle$ and its modulation amplitude $\Delta d = \mathrm{max}\, |d(x) -\langle d\rangle|$. Left: magnetic lattice. Right: ionic lattice.}
  \label{fig:charge_densities}
\end{figure}

One important effect of the lattices is that, even though the added modulation does not change the average chemical potential, the charge density is strongly impacted.  As shown in Fig.~\ref{fig:charge_densities}, both the average charge density $\langle d\rangle$ and the amplitude at which the charge density is modulated, $\Delta d =  \mathrm{max} |d(x) -\langle d\rangle|$, increase with both $\alpha_b$ and $\alpha_\mu$.  In particular, the ionic lattice induces a strong modulation of the charge density.

%%%%%%%%%%%%%%%%%%%%%%%%%%%%%%%%%%%%%%%%%%%%%%%%%%%%%%%%%%%%%%%%%%%%%%%%%%%%%%%%%%%%%%%%%%%%%%%%%%%%%%%%%%%%%%%%%%%%%%%%%%%%%%%%%%%%%%%%%%%%%%%%%%%%%%%%%%%%%%%%%%%%%%%%%%%%%%%%%%%%%%%%%%%%%%%%%%%%%%%%%%%%%%%%%%%%%%%%%%%%%%%%%%%%%%%%%%%%%%%%%%%%%%%%%%%%%%%%%%%%%%%%%%%%%%%%%%%%%%%%%%%%%%%%%%%%%%%%%%%%%%%%%%%%%%%%%%%%%%%%%%%%%%%%%%%%%%%%%%%%%%%%%%%%%%%%%%%%%%%%%%%%%%%%%%%%%%%%%%%%%%%%%%%%%%%%%%%%%%%%%%%%%%%%%%%%%%%%%%%%%%%%%%%%%%%%%%%%%%%%%%%%%%%%%%%%%%%%%%%%%%%%%%%%%%%%%%%%%%%%%%%%%%%%%%%%%%%%%%%%%%%%%%%%%%%%%%%%%%%%%%%%%%%%%%%%%%%%%%%%%%%%%%%%%%%%%%%%%%%%%%%%%%%%%%

\section{Conductivities}\label{sec:conductivities}

We now turn to our main topic, the linear electric conductivity of the striped state in the presence of an explicit lattice.  We first present formulas for the DC conductivities in terms of the horizon data of the background solution.  Then we present numerical computations of the optical conductivity, first for the magnetic lattice and then for the ionic case.

\subsection{DC conductivities}

We computed in \cite{Jokela:2016xuy} the DC conductivities of the striped solution in the absence of a lattice.  This computation involved the method developed in Refs.~\cite{Donos:2015gia,Banks:2015wha,Donos:2015bxe,Araujo:2015hna} using conserved bulk quantities to express the currents in terms of quantities which depend only on background fields at the horizon. To describe a system with spontaneous symmetry breaking, we generalized this method to include the translational Goldstone mode.  The stripes then slide at a velocity $v_s$ proportional to the applied electric field $E_x$.

Introducing a lattice does not alter the computation of \cite{Jokela:2016xuy} except that, because the Goldstone mode is lifted, the stripes can no longer slide.  The averaged\footnote{Spatial averages are denoted by $\langle\ldots\rangle=\int_0^L dx(\ldots)/L$.} DC conductivities are therefore given by nearly the same expressions as in \cite{Jokela:2016xuy}:
\begin{align} \label{sigmaxxDC}
 \langle \sigma_{xx}\rangle & =  \langle \hat\sigma^{-1}\rangle^{-1}  &\\ 
  & \, \  +\delta_{\alpha_b,\alpha_\mu,0}\,\frac{v_s}{E_x}\!\left[\sqrt 2\langle c(\psi_0)a'_{y,0}\rangle -\langle a_{t,0}\hat\sigma(a'^2_{y,0}\!+\psi'^2_0\!+z'^2_0)\rangle +\langle a_{t,0}\rangle\langle \hat\sigma^{-1}\rangle^{-1}\!-\langle a_{t,0}\hat\sigma\rangle\right] \nonumber \\
\label{sigmayyDC}
  \langle \sigma_{yy}\rangle & =  \left\langle \hat\sigma(1+z'^2_0+\psi'^2_0)+\frac{1}{\hat\sigma}\left(\sqrt 2 c(\psi_0)-\hat\sigma a_{t,0} a'_{y,0}\right)^2\right\rangle & \\
  \langle \sigma_{xy}\rangle & =  \langle \sigma_{yx}\rangle  = 0 \ , &
\end{align}
where
\bea
 \hat\sigma = \frac{\sqrt{(1+8\sin^4\psi_0(x))(1+8\cos^4\psi_0(x))}}{2\sqrt{2(1-a_{t,0}(x)^2)(1+a'_{y,0}(x)^2+\psi'_0(x)+z'^2_0(x))}} 
\eea
and the subscript $0$ denotes values of the background fields evaluated at the horizon, with the exception of $a_t(x,u)=a_{t,0}(x)(u-1)+{\cal O}\left((1-u)\right)$.\footnote{For $x$-dependent expressions, we refer the reader to Ref.~\cite{Jokela:2016xuy}.} 

We have introduced a Kronecker delta in the last term of \eqref{sigmaxxDC}, signaling an abrupt change of physics in the absence of explicit translation symmetry breaking $\alpha=0$. This discontinuity is a result of computing the conductivity to linear order.  Because the finite modulation $\alpha$ is parametrically larger than the infinitesimal electric field $E_x$, there is no way to overcome the pinning potential and cause the stripes to slide across the lattice.  We hope in future work to compute the nonlinear conductivity in response to finite $E_x$.

Although the addition of this Kronecker delta may seem ad hoc, we verify that it is correct in Secs.~\ref{sec:magnetic_lattice} and \ref{sec:ionic_lattice} by matching $\sigma_{xx}^\mathrm{DC}$ from Eq.~\eqref{sigmaxxDC} to the zero-frequency limit of the optical conductivity.

\subsection{Magnetic lattice}
\label{sec:magnetic_lattice}

We now focus our attention on the specific case of the magnetic lattice and impose the boundary condition \eqref{magnetic_boundary_condition} on $a_y$.  We construct the modulated background numerically as in Ref.~\cite{Jokela:2014dba}.  The DC conductivities can directly be computed from Eqs.~\eqref{sigmaxxDC} and \eqref{sigmayyDC} using the horizon values of the numerical solution.

To compute the optical conductivity, we consider linear fluctuations on top of the the striped backgrounds with the form:
\be 
 f = \bar f(x,u) + e^{-i\omega t} (1-u)^{-i\omega/4} \delta f(x,u) \ ,
\ee 
where $f$ represents each of the bulk fields $\psi$, $z$, $a_t$, $a_y$, and $a_x$. To turn on electric fields $e_x$ or $e_y$, we choose one of the following boundary conditions:
\begin{align} \label{Exbc}
 i \omega \delta a_x (x,0) + \partial_x \delta a_t(x,0) &= i \omega\, e_x & \\
 \delta a_y (x,0) &= e_y \ . & \label{Eybc} 
\end{align}
As usual, an extra factor of $\omega$ was included in the definitions of $e_x$ and $e_y$ so that the physical electric fields are $\propto i\omega\, e_{x,y} e^{-i\omega t}$. No other sources are turned on, so that
\be
 \partial_u \delta \psi(x,0) = 0 = \delta z(x,0) \ .
\ee
In addition, we require infalling conditions, i.e., that $\delta f$ are regular at the horizon, and that $\delta a_t(x,1) = 0$.   We then solve the linearized equations of motion numerically and extract the conductivities as follows:
\be
 \left(\begin{array}{c}
        j_x(\omega,x) \\ 
        j_y(\omega,x)
       \end{array}
\right)
 =  \left(\begin{array}{c}
        \partial_u \delta \hat a_x(x,u=0)\\ 
        \partial_u \delta \hat a_y(x,u=0)
       \end{array}
\right)
 =\  \left(\begin{array}{cc}
        \sigma_{xx}(\omega,x) & \sigma_{xy}(\omega,x) \\ 
        \sigma_{yx}(\omega,x) & \sigma_{yy}(\omega,x)
       \end{array}
\right)\left(\begin{array}{c}
        i\omega\,e_x\\ 
        i\omega\,e_y
       \end{array}\right) \ .
\ee
For further details, we refer the reader to \cite{Jokela:2016xuy}, and for a similar computation with complementary discussion, see \cite{Araujo:2015hna}.

\begin{figure}[!ht]
\center
 \includegraphics[width=0.50\textwidth]{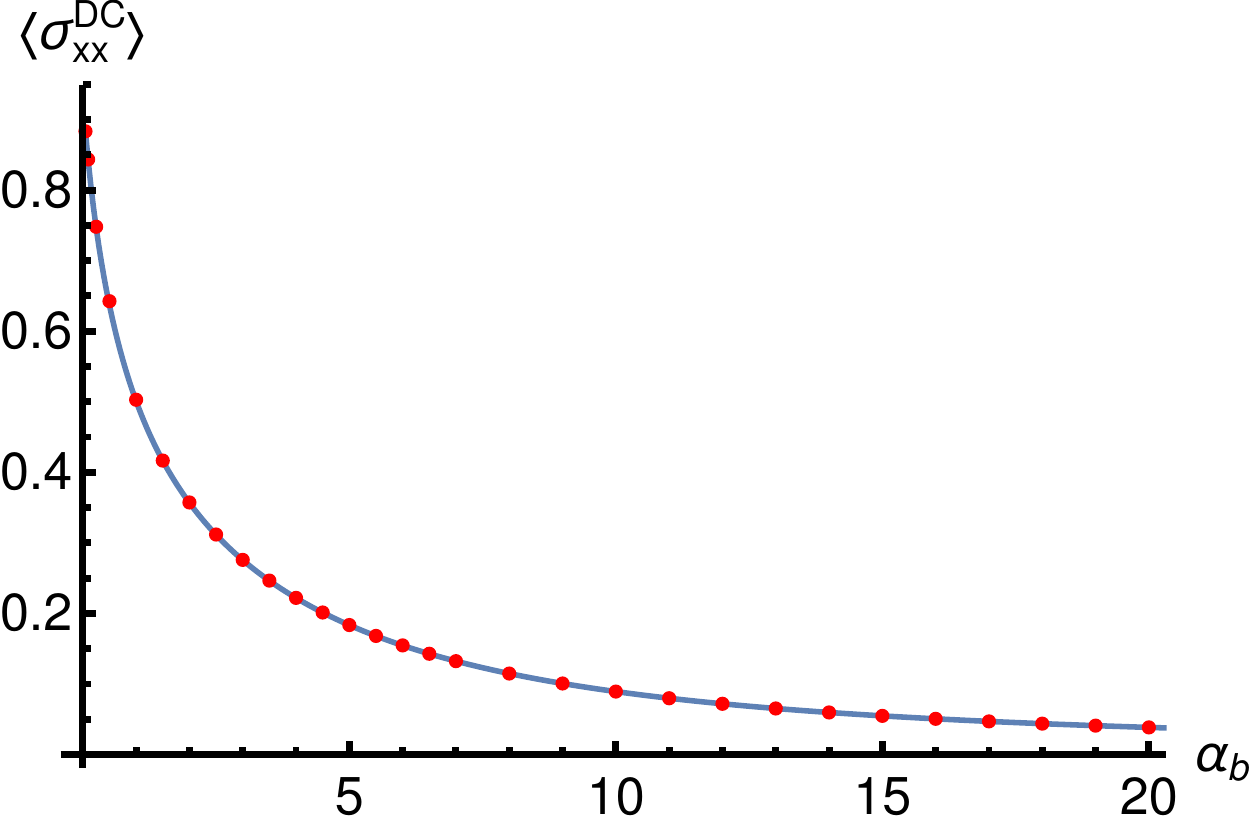}%
 \includegraphics[width=0.50\textwidth]{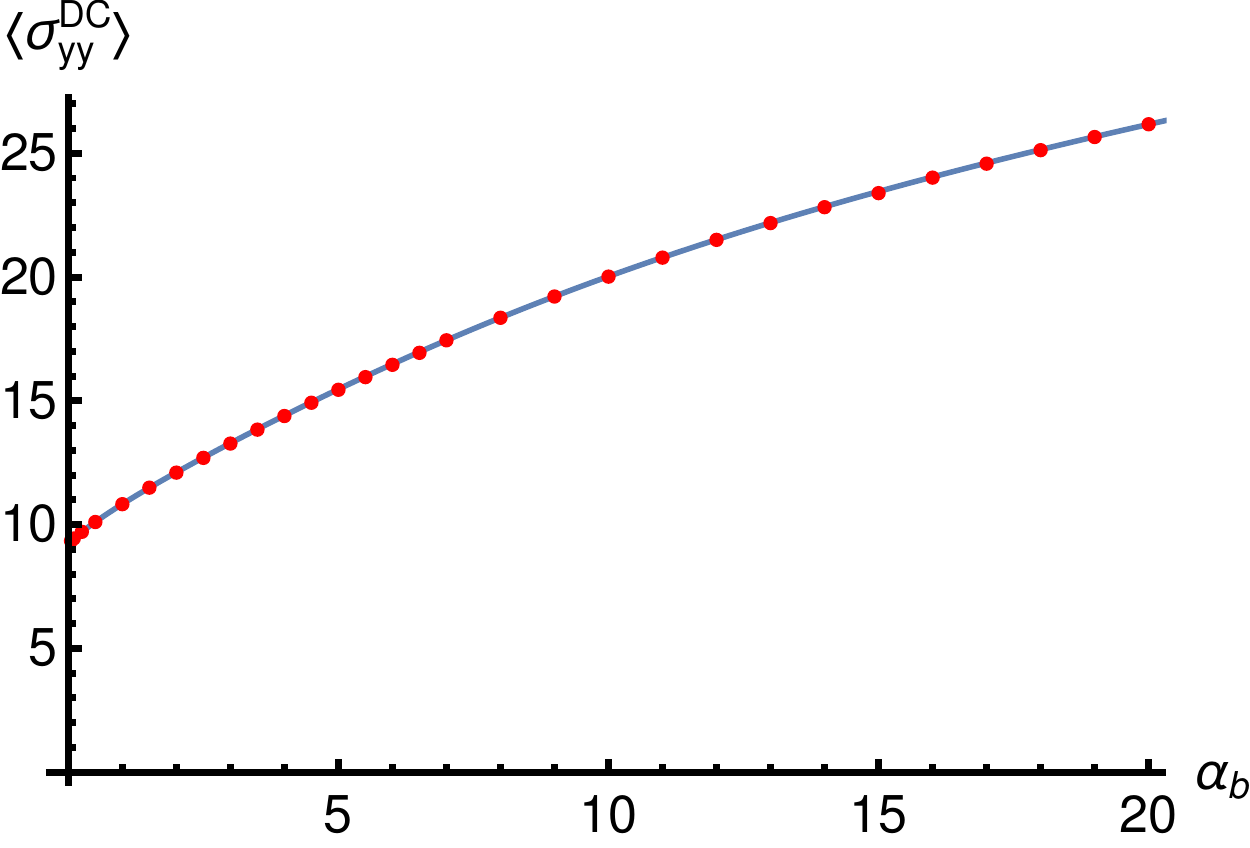}
 \caption{Comparison of the numerical $\omega \to 0$ limit of the optical conductivities (red dots) to the formula~\protect\eqref{sigmaxxDC} for the DC conductivities (blue curves), plotted against the amplitude of the magnetic lattice $\alpha_b$. Left: $\langle\sigma^\mathrm{DC}_{xx}\rangle$. Right:  $\langle\sigma^\mathrm{DC}_{yy}\rangle$.}
  \label{fig:DCcomparison_magnetic}
\end{figure}

Our first goal is to match the $\omega \to 0$ limit of the optical conductivities computed numerically with the DC conductivities computed from Eqs.~\eqref{sigmaxxDC} and \eqref{sigmayyDC}.  We plot these in Fig.~\ref{fig:DCcomparison_magnetic}, and they match to excellent accuracy.\footnote{The zero-frequency limit of the optical conductivity (red dots in Fig.~\ref{fig:DCcomparison_magnetic}) is extracted as follows. For conductivity in the $x$ direction, the dependence of the $\langle \sigma_{xx} \rangle$ on $\omega$ was so weak that we simply take the data at lowest available $\omega$. For $\langle \sigma_{yy} \rangle$ in $y$ direction, we fit the data at low $\omega$ to a Drude form (see Fig.~\ref{fig:sigma_yyB} below) because the peak at small $\omega$ becomes so narrow, in particular at large $\alpha_b$, that extrapolation to $\omega = 0$ is needed.}

\begin{figure}[!ht]
\center
 \includegraphics[width=0.53\textwidth]{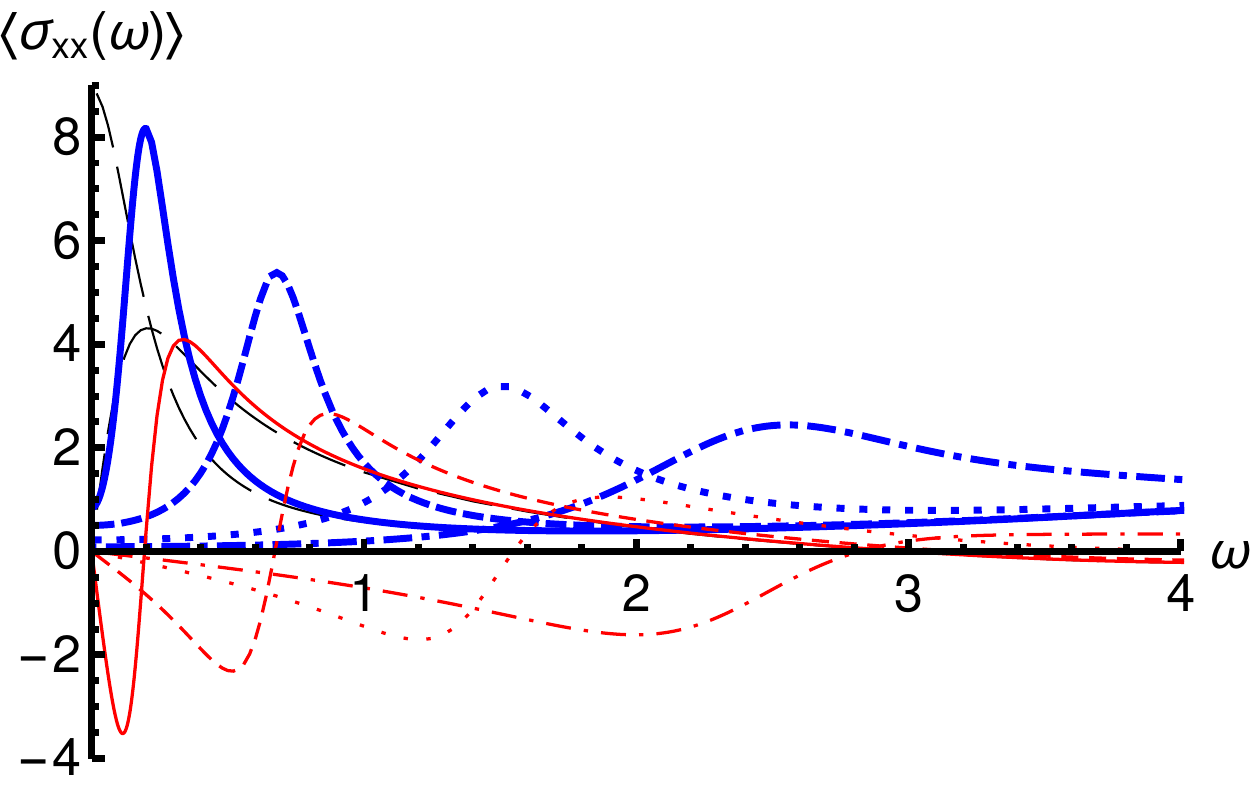}%
 \includegraphics[width=0.47\textwidth]{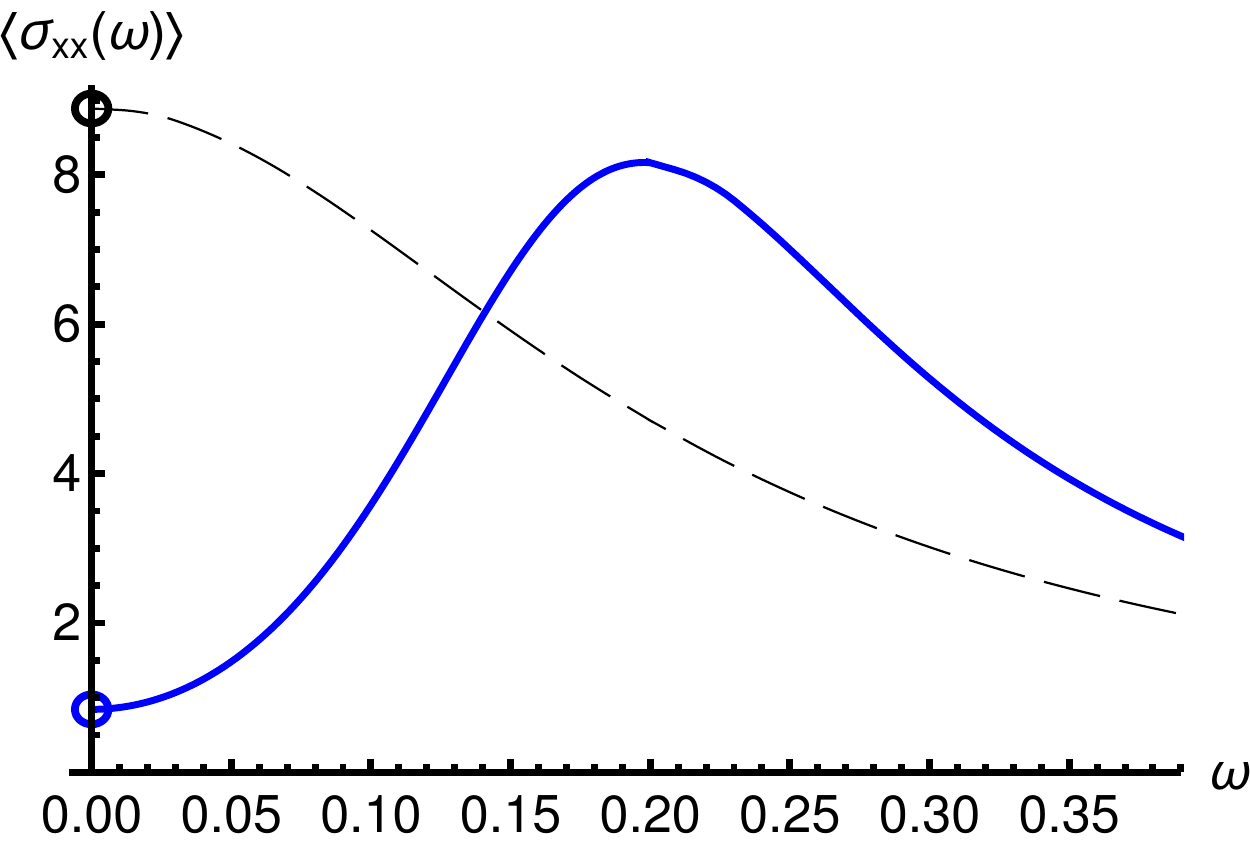}
 \caption{Left: The averaged optical conductivity $\langle\sigma_{xx}\rangle$ for a magnetic lattice.  The real part is shown in thick, blue curves and imaginary part in thin, red curves. Solid, dashed, dotted, and dot-dashed curves correspond to $\alpha_b=0.1$, 1, 4, and 10, respectively.  For comparison, we also show the result at $\alpha_b=0$ as thin, long-dashed black curves. Right: A zoom into the region with low $\omega$, showing the optical conductivities at $\alpha_b=0$ and at $\alpha_b=0.1$. The circles are the values of the DC conductivities from Eq.~\protect\eqref{sigmaxxDC} (The higher value includes the contribution from the sliding stripes~\protect\cite{Jokela:2016xuy}).} 
  \label{fig:sigma_xxB}
\end{figure}

Turning now to nonzero $\omega$, our results for the optical conductivity $\langle \sigma_{xx}(\omega) \rangle$ for the magnetic lattice are shown for various values of $\alpha_b$ in Fig.~\ref{fig:sigma_xxB} (left).   As $\alpha_b$ increases, the peak in ${\rm Re} \langle \sigma_{xx}(\omega) \rangle$, located at $\omega=0$ for $\alpha_b=0$, shifts to higher frequencies.  In addition, the height of the peak shrinks and the width broadens.  Notably, the conductivity at small $\omega$ immediately drops by an order of magnitude when $\alpha_b$ becomes nonzero, as is demonstrated more clearly in the right hand plot of Fig.~\ref{fig:sigma_xxB}.

This result can be fit very well with a Drude-Lorentz model of the following form:
\be \label{sigmaxxform}
 \langle\sigma_{xx}(\omega)\rangle = \frac{\langle\sigma_{xx}^\mathrm{DC}\rangle}{1-i\tau_{xx} \omega} + \frac{K_{xx}\tau_{xx}}{1-i\tau_{xx} \omega\left(1-\frac{\omega_{xx}^2}{\omega^2}\right)}
  = \frac{\langle\sigma_{xx}^\mathrm{DC}\rangle}{1-i\tau_{xx} \omega} + \frac{ i K_{xx}\omega}{\omega^2-\omega_{xx}^2 +i\omega/\tau_{xx}} \ .
\ee

\begin{figure}[!ht]
\center
 \includegraphics[width=0.50\textwidth]{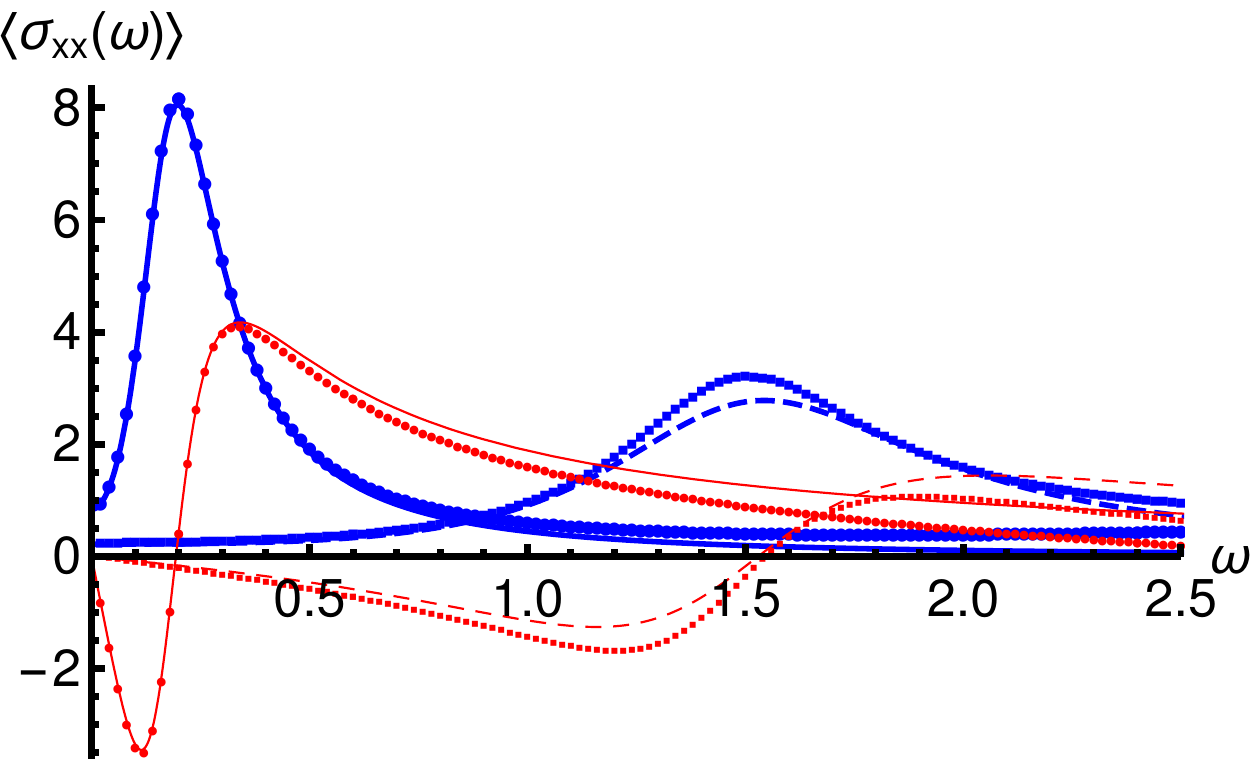}% 
 \includegraphics[width=0.50\textwidth]{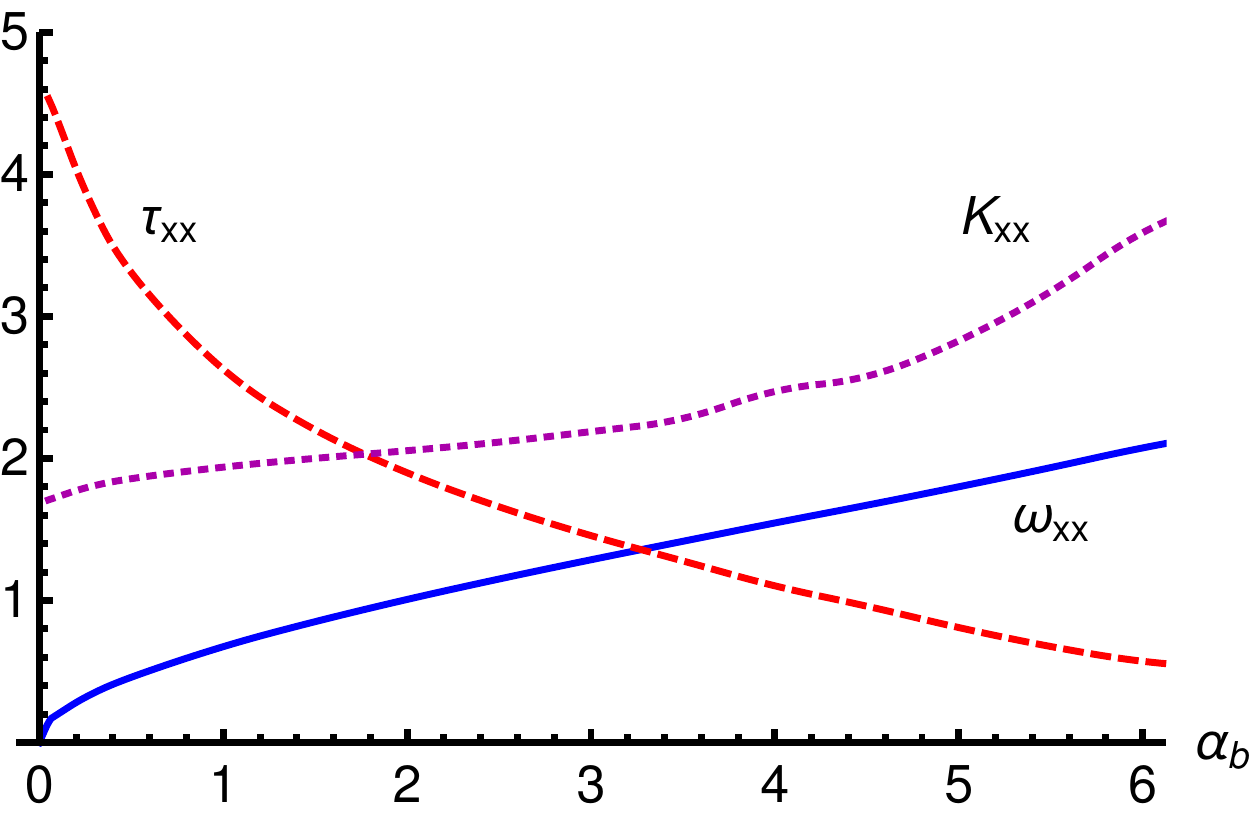}
 \caption{The averaged optical conductivity $\langle\sigma_{xx}(\omega)\rangle$. Left: The dots show the numerical data, and the curves are the results of the fit. Real parts are the thick, blue  curves and larger dots, and imaginary parts are the thin, red curves and smaller dots. Solid curves and round dots are at $\alpha_b=0.1$,  whereas dashed curves and boxes are for $\alpha_b=4$. Right: The three fit parameters $K_{xx}$, $\tau_{xx}$, and $\omega_{xx}$ as functions of $\alpha_b$.} 
 \label{fig:sigma_xxB_fit}
\end{figure}

There are three parameters $K_{xx}$, $\tau_{xx}$, and $\omega_{xx}$ which we fit to the data,\footnote{The fit for $\sigma_{xx}$ (as well as the fit for $\sigma_{yx}$ below) was done using a least-squares method, using the data within a range of about two half-widths around the peak.} and the results are plotted in Fig.~\ref{fig:sigma_xxB_fit}.  In the left hand plot we show examples of the fit compared to the data at two values for the amplitude of modulation, $\alpha_b = 0.1$ and $\alpha_b = 4$. 
The conductivity \eqref{sigmaxxform} is a sum of two terms.  The first term is the Drude form and describes the residual metallicity of charge carriers flowing across the stripes.  Since we already demonstrated above that the zero-frequency limit of the  optical conductivity matches with the result of Eq.~\eqref{sigmaxxDC}, we fix $\langle\sigma_{xx}^\mathrm{DC}\rangle$ by using this formula.

The second term is a Lorentzian which describes pinned stripes \cite{Lee, Gruner} and results from modeling the motion of the stripes in the lattice potential as a driven, damped harmonic oscillator.  The resonance frequency $\omega_{xx}$ is related to the lattice potential.  The harmonic oscillator model predicts $\omega_{xx} \sim \alpha_b^{1/2}$, which roughly fits the data for small $\alpha_b$.  

Notice that the second term vanishes as $\omega \to 0$, and the stripes do not contribute to the DC conductivity.  The exception is when $\omega_{xx} = 0$, in which case there is an extra contribution to the DC conductivity is given by $K_{xx} \tau_{xx}$. In fact, $\omega_{xx}$ vanishes precisely is the absence of the lattice potential.  Then the stripes can slide \cite{Jokela:2016xuy}, and the result agrees with the DC formula~\eqref{sigmaxxDC}, which similarly has an extra term which is nonzero when $\alpha_b=0$.

The fit of the data to Eq.~\eqref{sigmaxxform} is generally quite good. Notice that we simultaneously fit both the real part and the imaginary part of the conductivity with the same parameter values. At small $\alpha_b$, it is best for $\omega \lesssim 1$. The fit is worse at larger $\omega$ because there is a ``continuum'' contribution to the conductivity which is not captured by the formula~\eqref{sigmaxxform}. The quality of the fit also drops with increasing $\alpha_b$, as the peak moves to higher frequencies and interferes with the continuum part. For $\alpha_b \gtrsim 6$ (not shown) the fit fails to reproduce the $\omega$-dependence of the data.

We made the assumption in Eq.~\eqref{sigmaxxform} that the decay times of the Drude contribution and the Lorentzian are the same, which does not necessarily need to be the case. However, recall that the DC conductivity of the pinned system is highly suppressed, so consequently the contribution from Drude peak is subleading by roughly an order of magnitude with respect to the Lorentzian. Therefore, our fit is not sensitive to the details of the Drude peak, and because of this, we have chosen not to fit its decay time independently. Instead, we have tested that replacing the first term in~\eqref{sigmaxxform} by a different formula, e.g., a constant~\cite{Delacretaz:2016ivq} does not improve the fit significantly. If we had an access to the quasinormal mode spectrum, we could systematically include higher order poles contributing to the optical conductivities \cite{Amado:2008ji}.

We now turn to the Hall conductivity $\sigma_{yx}$. As discussed in Sec.~\ref{sec:spontaneous_stripes}, in the absence of a lattice, $\sigma_{yx}$ contains a delta peak at $\omega=0$.  As the stripes slide due to an infinitesimal electric field, the finite persistent transverse current at any fixed location varies, leading to an infinite DC Hall conductivity \cite{Jokela:2016xuy}.

\begin{figure}[!ht]
\center
 \includegraphics[width=0.70\textwidth]{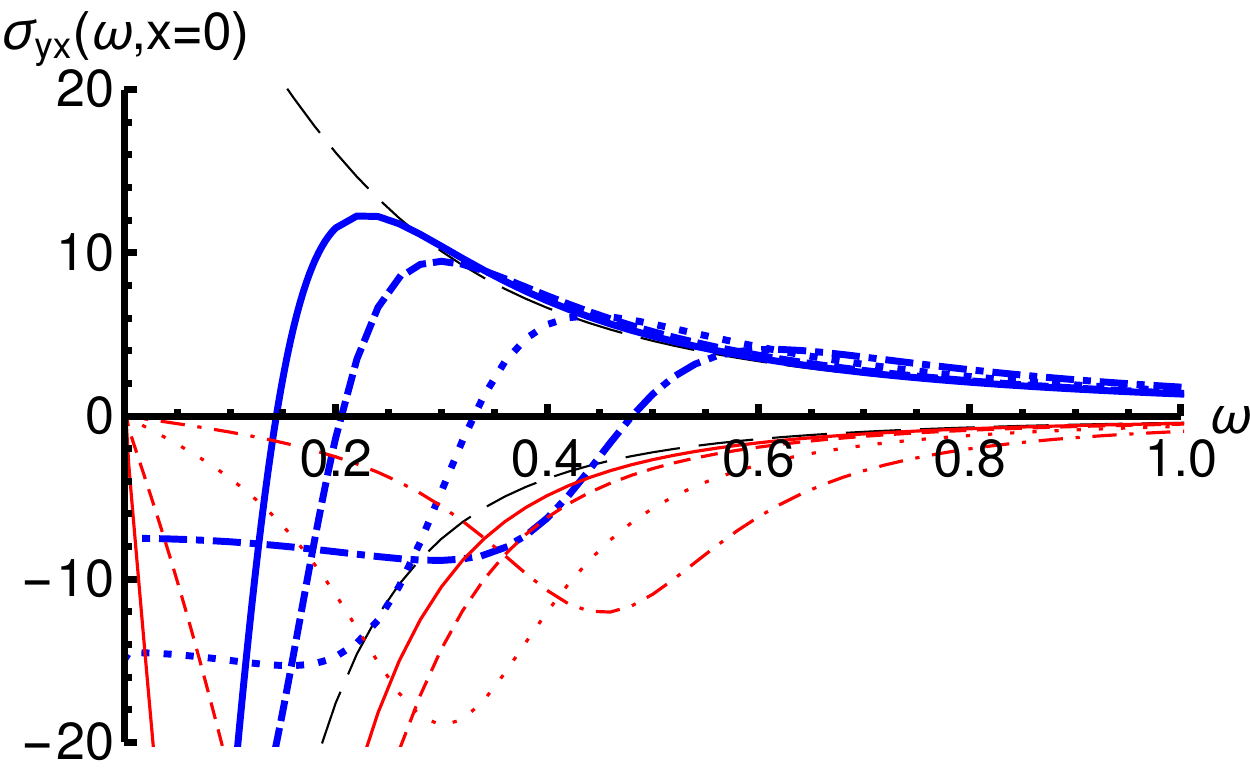}
 \caption{The optical conductivity $\sigma_{yx}(\omega)$ at $x=0$ for small $\omega$ and small lattice amplitude $\alpha_b$.  The real parts are denoted by thick, blue  curves, and the imaginary parts are shown by thin, red curves. Solid, dashed, dotted, and dotdashed curves correspond to $\alpha_b=0.05$, $0.1$, $0.25$, and $0.5$, respectively, and the results for $\alpha_b=0$ are shown as thin, long-dashed, black curves.} 
  \label{fig:sigma_yxB}
\end{figure}

As we turn on nonzero $\alpha_b$, the delta peak is regulated, becoming lower, broader, and moving to larger $\omega$. Our numerical results are shown in Fig.~\ref{fig:sigma_yxB}.  Although the Hall conductivity is nonzero, its spatial average vanishes. So, instead of $\langle \sigma_{yx} \rangle$, we plot the $\sigma_{yx}$ at a specific point, $x=0$. And, the result oscillates as $x$ is varied.

\begin{figure}[!ht]
\center
 \includegraphics[width=0.50\textwidth]{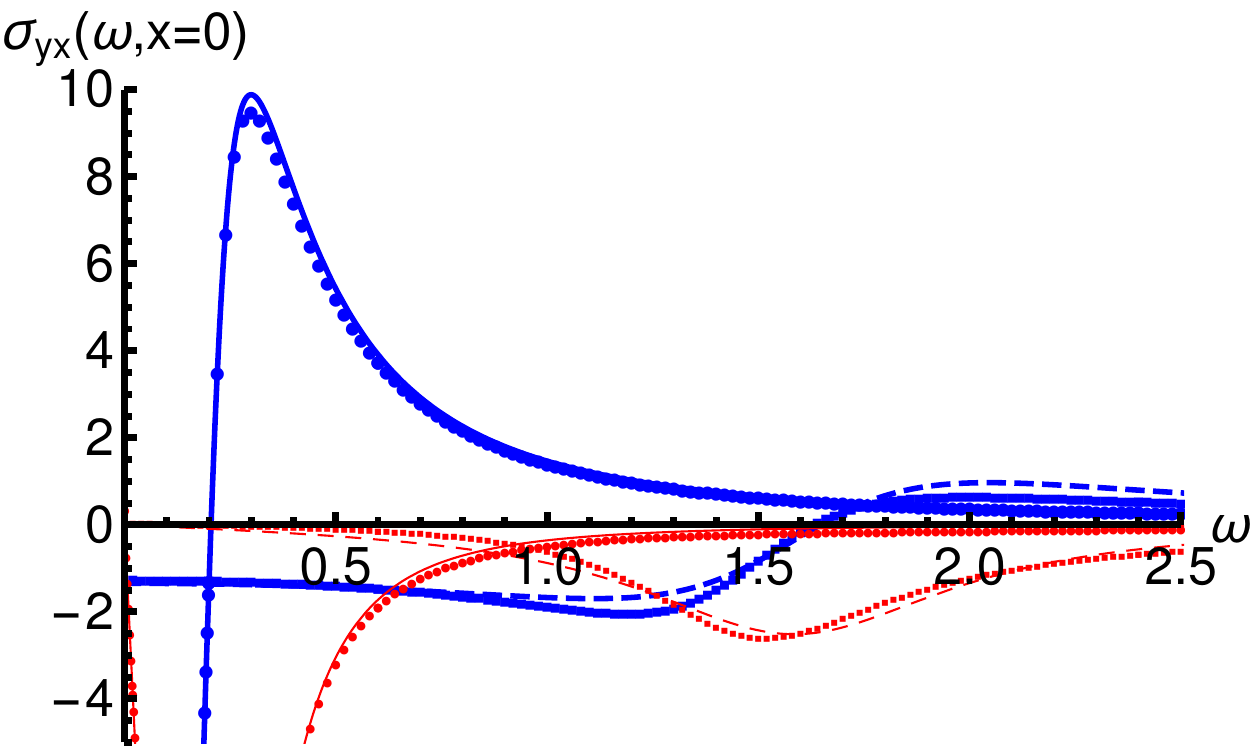}% 
 \includegraphics[width=0.50\textwidth]{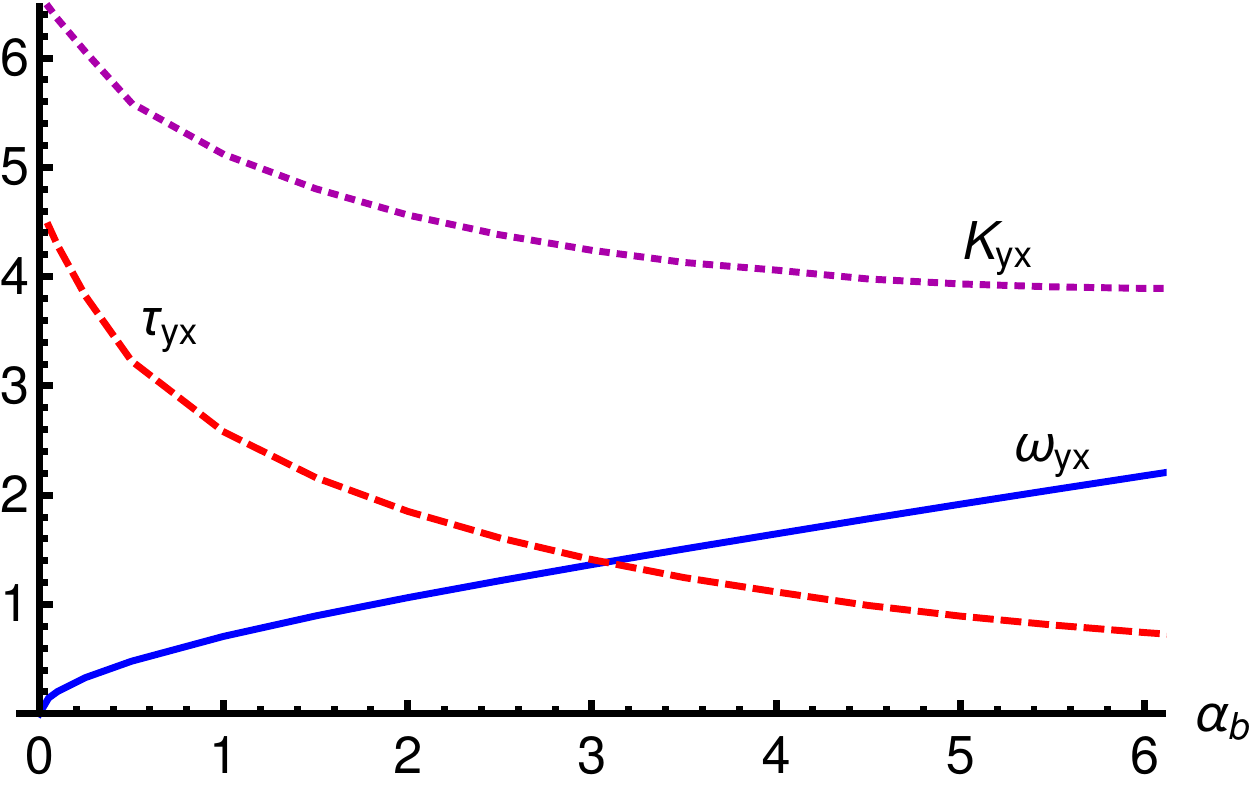}
 \caption{The Hall conductivity $\sigma_{yx}(\omega)$ at $x=0$. Left: The dots show the numerical data, and the curves are the results of the fit. Real parts are the thick, blue curves and larger dots, and imaginary parts are the thin, red curves and smaller dots. Solid curves and round dots show $\alpha_b=0.1$, whereas dashed curves and boxes shown $\alpha_b=4$. Right: The three fit parameters $K_{yx}$, $\tau_{yx}$, and $\omega_{yx}$ as functions of $\alpha_b$.} 
  \label{fig:sigma_yxB_fit}
\end{figure}

We fit the Hall conductivity at nonzero $\alpha_b$ to the following modified Lorentzian form:\footnote{Notice that the parameters $\omega_{yx}$ and $\tau_{yx}$, which determine the location of the resonance on the complex $\omega$-plane, do not depend on $x$.}
\be \label{sigmayxform}
 \sigma_{yx}(\omega,x) = \frac{ K_{yx}(x)/\tau_{yx}}{\omega^2-\omega_{yx}^2 +i\omega/\tau_{yx}} \ ,
\ee
with three parameters $K_{yx}$,  $\omega_{yx}$, and $\tau_{yx}$ which we fit to the numerical data.  The results are shown in Fig.~\ref{fig:sigma_yxB_fit}.  
As we saw for $\sigma_{xx}$ above, the fit is very good at small $\alpha_b$ but deteriorates as $\alpha_b$ grows. All the $x$ dependence is in $K_{yx}$, which to good accuracy varies as $\cos(2 \pi x/L)$.  We observe that, to within the precision of the fits, $\omega_{xx} = \omega_{yx}$ and $\tau_{xx} = \tau_{yx}$, which is expected since the peaks in the two components of the conductivity are due to the same resonant physics.

The modified Lorentzian form of \eqref{sigmayxform} comes from the same driven, damped harmonic oscillator model of the stripes as in \eqref{sigmaxxform}.  The difference between \eqref{sigmayxform} and \eqref{sigmaxxform} can be understood by their different relationship to the motion of the stripes.  The Hall current depends on the location of the stripes, in particular, the local value of the persistent current, while the longitudinal current depends on the velocity of the stripes.  The difference in the conductivities then amounts to an extra time derivative in $\sigma_{xx}$, which yields an extra factor of $i\omega$ in \eqref{sigmaxxform} compared to \eqref{sigmayxform}.

Moreover, notice that as $\omega_{yx} \to 0$, the modified Lorentzian \eqref{sigmayxform} can be written as a sum of a delta peak and a Drude form,
\be \label{deltapeakalpha0}
\sigma_{yx}(\omega,x,\alpha=0) = \frac{\tau_{yx} K_{yx}(x) }{1-i\tau_{yx} \omega} - K_{yx}(x)\left(\frac{i }{\omega}+\pi\, \delta(\omega)\right) \ ,
\ee
which correctly describes the Hall conductivity at small $\omega$ in the absence of pinning~\cite{Jokela:2016xuy}. This result requires that the modified Lorentzian~\eqref{sigmayxform} is constant at small $\omega$, which is not the case for the Lorentzian form in~\eqref{sigmaxxform}.

\begin{figure}[!ht]
\center
 \includegraphics[width=0.50\textwidth]{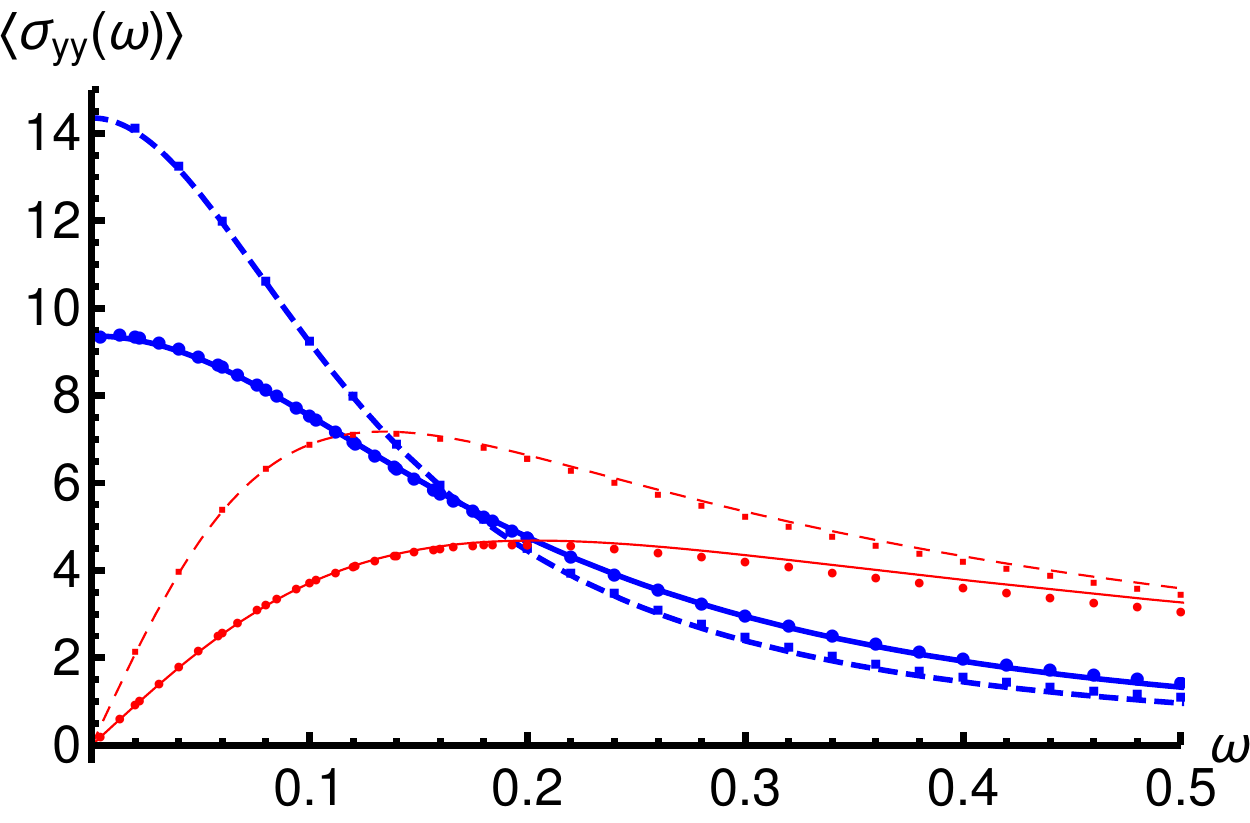}% 
 \includegraphics[width=0.50\textwidth]{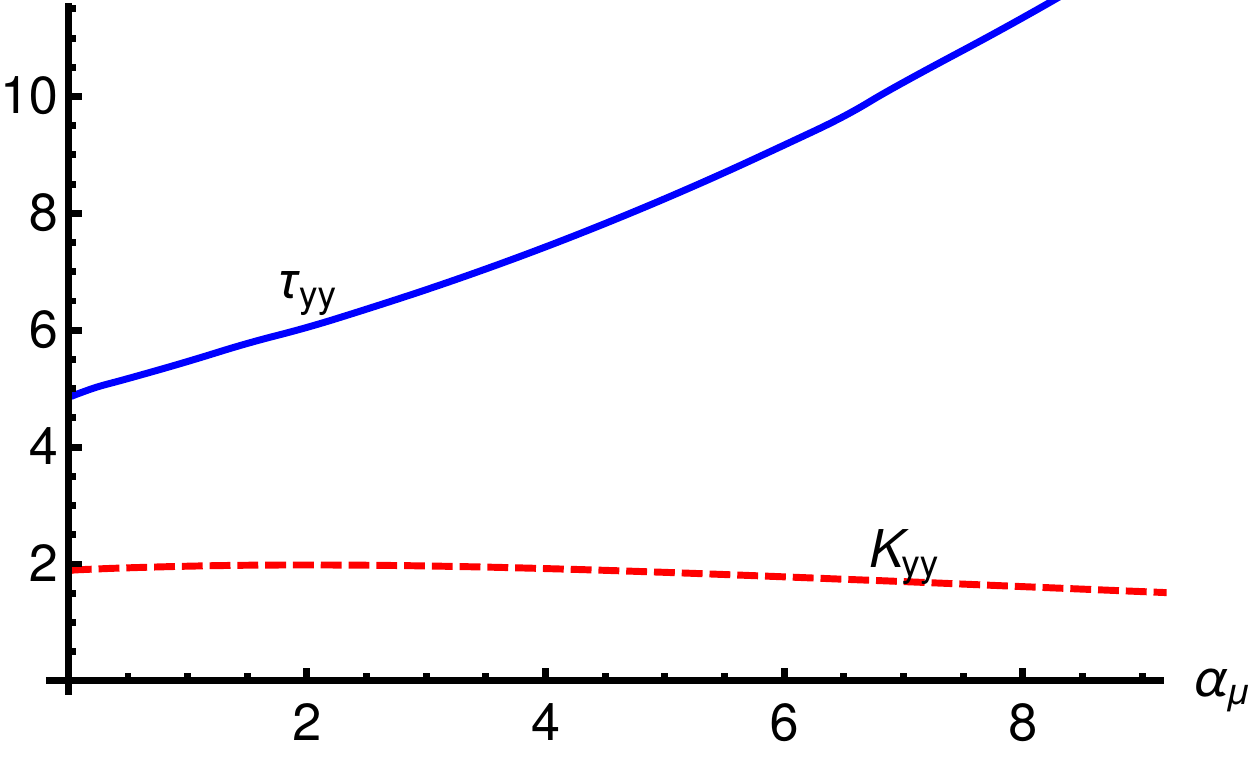}
 \caption{The averaged optical conductivity $\langle\sigma_{yy}(\omega)\rangle$. Left: The dots show the numerical data, and the curves are the results of the fit. Real parts are the thick, blue  curves and larger dots, and imaginary parts are the thin, red curves and smaller dots. Solid curves and round dots are at $\alpha_b=0.1$,  whereas dashed curves and boxes are for $\alpha_b=4$. Right:  The fit parameters $\tau_{yy}$ and $K_{yy}$ as functions of $\alpha_b$.} 
  \label{fig:sigma_yyB}
\end{figure}

Finally, we consider the conductivity parallel to the stripes.  Because translation invariance in the $y$ direction is preserved, we do not expect to see any pinning effects in $\sigma_{yy}$. Indeed, we observe instead that the conductivity, is well described in terms of a single Drude peak, 
\be 
\label{sigma_yy}
\langle \sigma_{yy} \rangle  = \frac{\langle \sigma_{yy}^\mathrm{DC} \rangle}{1-i\tau_{yy} \omega} = \frac{\tau_{yy}K_{yy}}{1-i\tau_{yy} \omega} \ .
\ee
The results of fitting the data to this form are shown in Fig.~\ref{fig:sigma_yyB}. The width of the Drude peak decreases and the DC conductivity increases with $\alpha_b$, so that the area of the peak (which is proportional to $K_{yy}$) stays roughly constant. The increase in the conductivity with $\alpha_b$ is in accordance with the increase in charge density demonstrated in Fig.~\ref{fig:charge_densities} (left). We do not, however, find a direct proportionality between the two observables, which might be due to a nonlinear contribution from the induced charge density related to the enhanced amplitude of the stripes.

%%%%%%%%%%%%%%%%%%%%%%%%%%%%%%%%%%%%%%%%%%%%%%%%%%%%%%%%%%%%%%%%%%%%%%%%%%%%%%%%%%%%%%%%%%%%%%%%%%%%%%%%%%%%%%%%%%%%%%%%%%%%%%%%%%%%%%%%%%%%%%%%%%%%%%%%%%%%%%%%%%%%%%%%%%%%%%%%%%%%%%%%%%%%%%%%%%%%%%%%%%%%%%%%%%%%%%%%%%%%%%%%%%%%%%%%%%%%%%%%%%%%%%%%%%%%%%%%%%%%%%%%%%%%%%%%%%%%%%%%%%%%%%%%%%%%%%%%%%%%%%%%%%%%%%%%%%%%%%%%%%%%%%%%%%%%%%%%%%%%%%%%%%%%%%%%%%%%%%%%%%%%%%%%%%%%%%%%%%%%%%%%%%%%%%%%%%%%%%%%%%%%%%%%%%%%%%%%%%%%%%%%%%%%%%%%%%%%%%%%%%%%%%%%%%%%%%%%%%%%%%%%%%%%%%%%%%%%%%%%%%%%%%%%%%%%%%%%%%%%%%%%%%%%%%%%%%%%%%%%%%%%%%%%%%%%%%%%%%%%%%%%%%%%%%%%%%%%%%%%%%%%%%%%%%

\subsection{Ionic lattice}
\label{sec:ionic_lattice}

We now replace the magnetic lattice with an ionic lattice by imposing the spatially modulated boundary condition \eqref{ionic_boundary_condition} on $a_t$. We repeat the numerical construction of the background and analysis of the fluctuations, as discussed in Sec.~\ref{sec:magnetic_lattice}. Many of the effects of the ionic lattice are qualitatively similar to the magnetic lattice, but we will highlight several relevant differences.

\begin{figure}[!ht]
\center
 \includegraphics[width=0.50\textwidth]{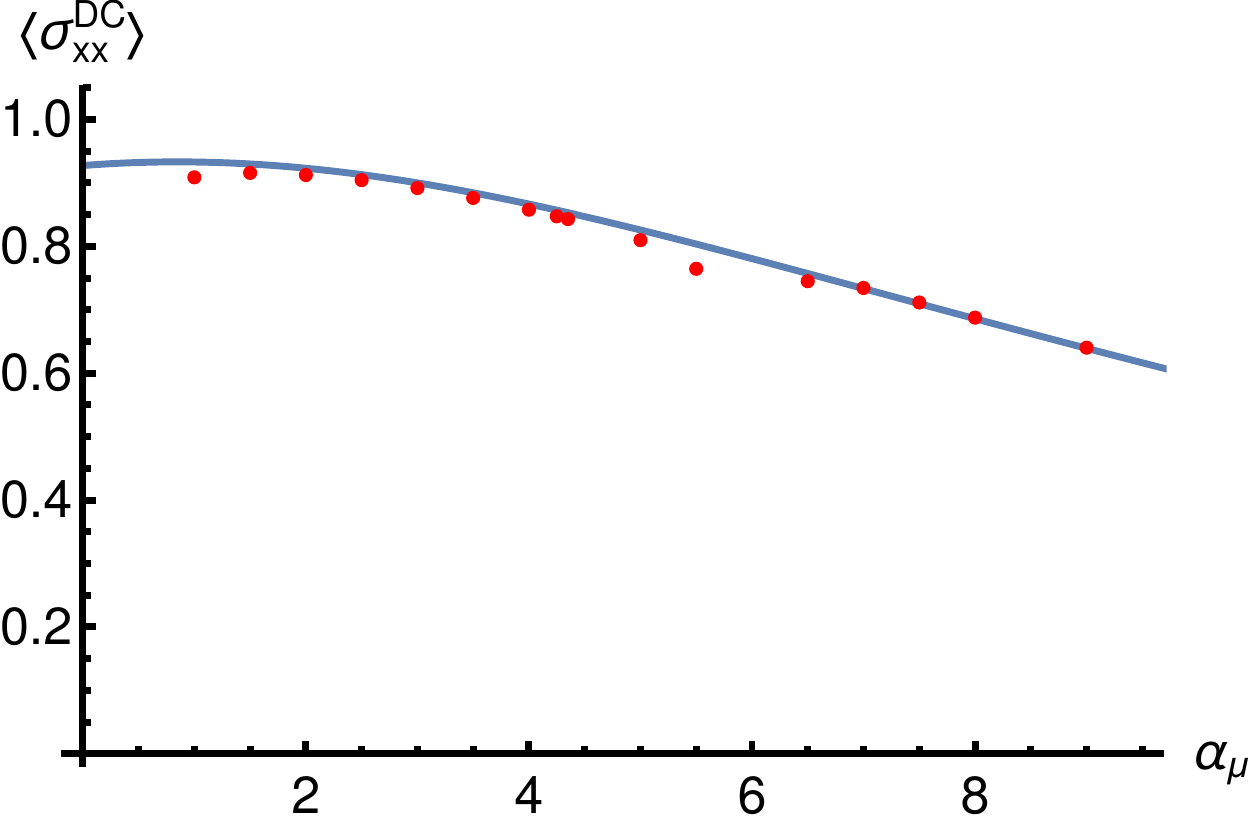}%
 \includegraphics[width=0.50\textwidth]{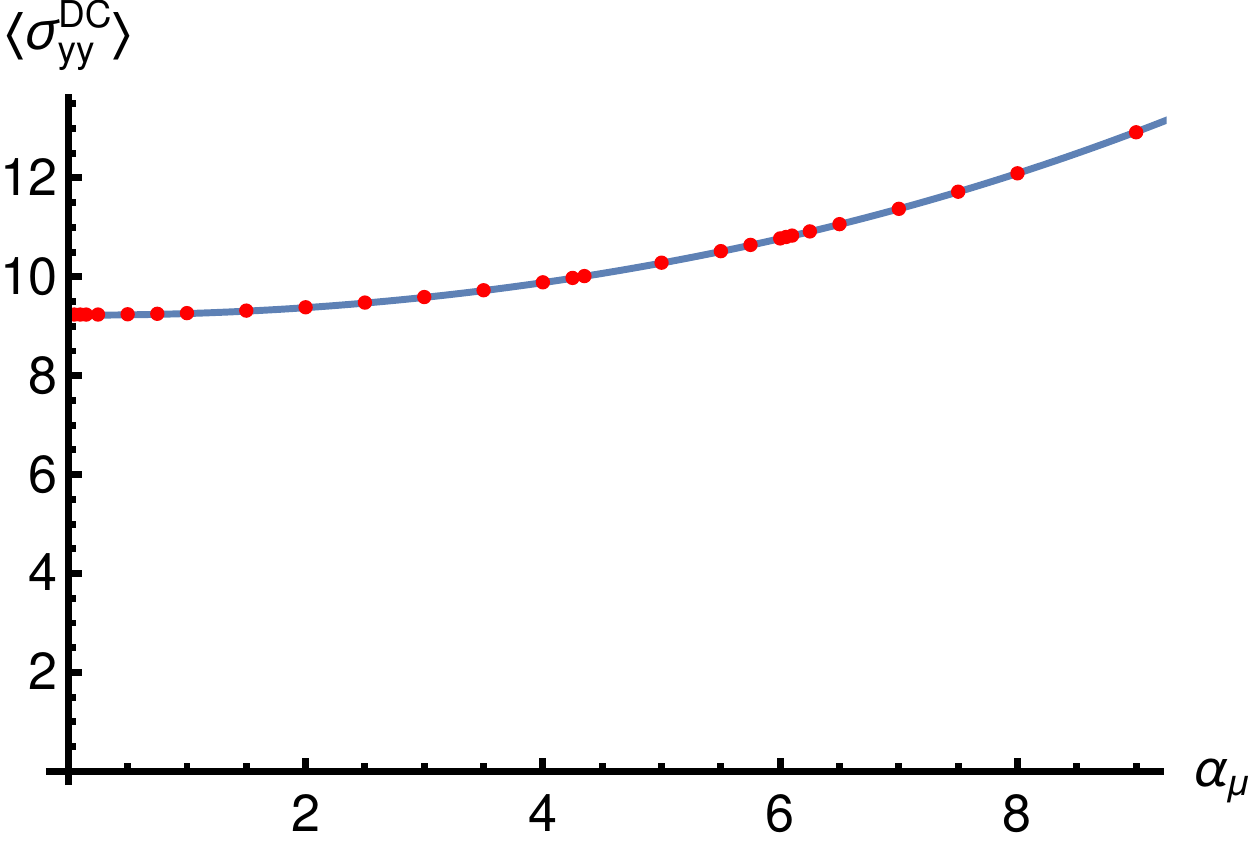}
 \caption{Comparison of numerical data for the $\omega \to 0$ limit of the optical conductivities (red dots) to the formulas~\protect\eqref{sigmaxxDC} and \protect\eqref{sigmayyDC} for the DC conductivities (blue curves). Left:   $\langle\sigma^\mathrm{DC}_{xx}\rangle$. Right: $\langle\sigma^\mathrm{DC}_{yy}\rangle$.}
  \label{fig:DCcomparison_ionic}
\end{figure}

\begin{figure}[!ht]
\center
 \includegraphics[width=0.50\textwidth]{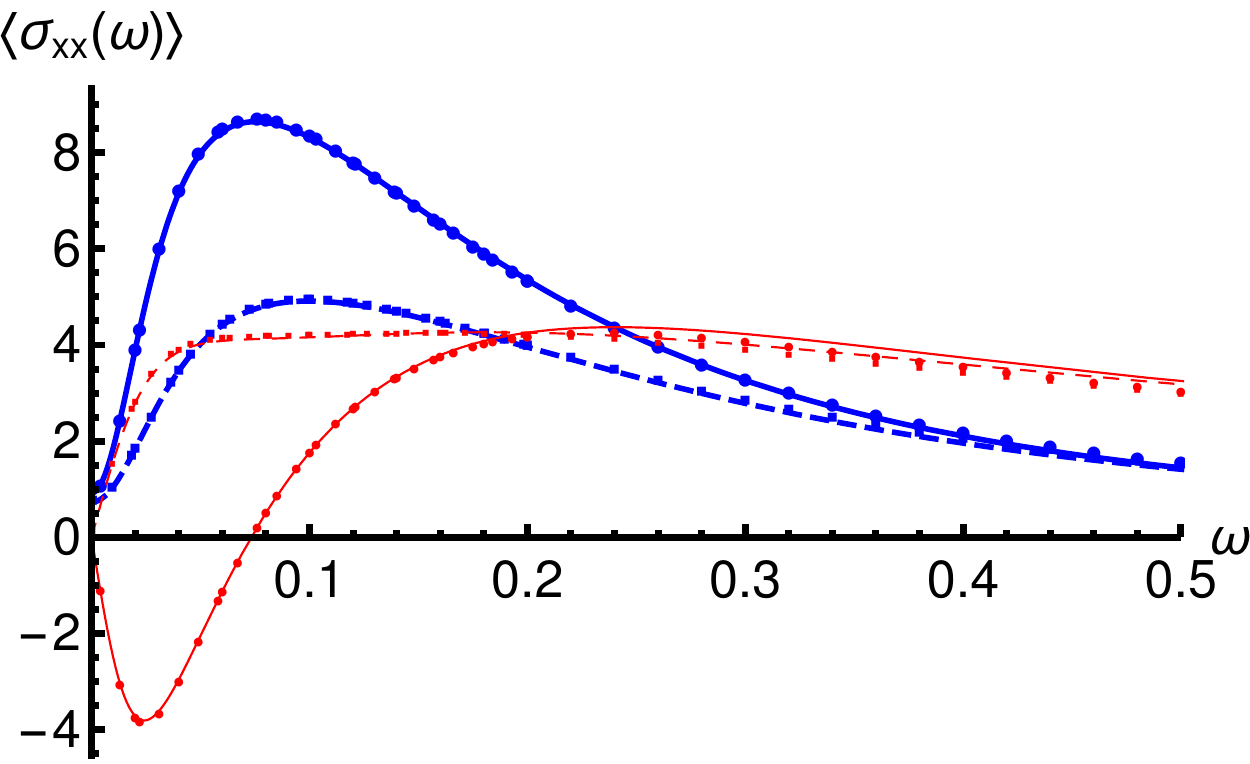}% 
 \includegraphics[width=0.50\textwidth]{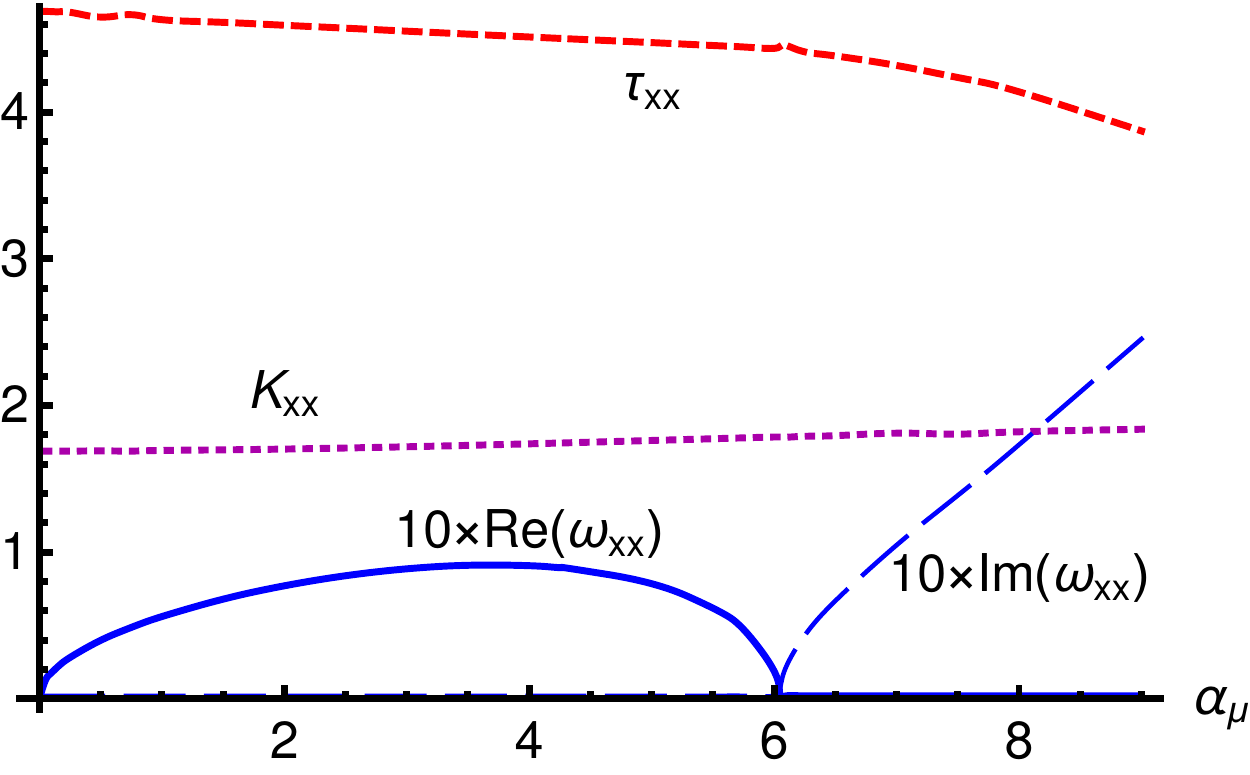}
 \caption{Fit results for the averaged optical conductivity $\langle\sigma_{xx}(\omega)\rangle$ with ionic lattice. Left: Numerical data (dots) compared to fit results (curves). Real parts are the thick, blue curves and larger dots, and imaginary parts are the thin, red curves and smaller dots. Solid curves and round dots are at $\alpha_\mu=2$, whereas dashed curves and boxes are for $\alpha_\mu=7$. Right: The three fit parameters $K_{xx}$, $\tau_{xx}$, and $\omega_{xx}$ as functions of $\alpha_\mu$.  The resonance frequency $\omega_{xx}$ is real for $\alpha_{\mu} \lesssim 6$, and the real part is shown as a solid curve.  For $\alpha_{\mu} \gtrsim 6$, it is purely imaginary, and the imaginary part is plotted with a dashed curve. In addition, the value of $\omega_{xx}$ has been multiplied by 10 to make it visible on the same scale as the other parameters.} 
  \label{fig:sigma_xxmu_fit}
\end{figure}

As with the magnetic lattice, the zero-frequency limit of the optical conductivity matches the DC computation using Eqs.~\eqref{sigmaxxDC} and \eqref{sigmayyDC}, as shown in Fig.~\ref{fig:DCcomparison_ionic}.  For $\langle\sigma_{xx}^\mathrm{DC}\rangle$, the slight mismatch is due to numerical error in the fitting of $\langle\sigma_{xx}(\omega)\rangle$ at small $\omega$.  The sharp peak in $\langle\sigma_{xx}(\omega)\rangle$ at small $\omega$, which is evident in Fig.~\ref{fig:sigma_xxmu_fit}, makes fitting the $\omega \to 0$ limit challenging. We fitted a polynomial to the data (for $\omega \ll 1$) in order to extrapolate to $\omega = 0$.

The optical conductivity $\sigma_{xx}^\mathrm{DC}$ perpendicular to the stripes can be analyzed as in the case of the magnetic lattice.  We again fit the data\footnote{As the resonant peaks lie at low $\omega$, we choose the data points with $\omega < 0.5$ for the least-squares fits.} for the averaged conductivity to a combination of a Drude peak and a Lorentzian form
\be \label{sigmaxxform2}
 \langle\sigma_{xx}(\omega)\rangle
  = \frac{\langle\sigma_{xx}^\mathrm{DC}\rangle}{1-i\tau_{xx} \omega} + \frac{ i K_{xx}\omega}{\omega^2-\omega_{xx}^2 +i\omega/\tau_{xx}} \ .
\ee
In particular, since Fig.~\ref{fig:DCcomparison_ionic} shows that the $\omega \to 0$ limit of the conductivity agrees with the analytic expression~\eqref{sigmaxxDC}, we fix the coefficient of the Drude term by using this formula, as we did for the magnetic lattice. 
The results are shown in Fig.~\ref{fig:sigma_xxmu_fit}. Thanks to the small size of $\omega_{xx}$, the quality of the fit is clearly better than in the case of magnetic lattice, as one can see by comparing the left hand plots in Fig.~\ref{fig:sigma_xxB_fit} and Fig.~\ref{fig:sigma_xxmu_fit}.

However, there are some key differences between these results and the magnetic lattice results of Sec.~\ref{sec:magnetic_lattice}. First, the magnitude of the pinning frequency $\omega_{xx}$ is suppressed by an order of magnitude with respect to Fig.~\ref{fig:sigma_xxB_fit}. In Fig.~\ref{fig:sigma_xxmu_fit} (right), we multiplied the result by a factor of $10$ in order to make its structure visible. This is consistent with subleading charge modulation of the striped phase; explicit breaking in $a_t$ only weakly pins the stripes because the leading modulation is in $a_y$ and $\psi$ and the modulation of $a_t$ is two orders of magnitude smaller. 

Second, $\omega_{xx}$ hits zero and becomes imaginary for $\alpha_\mu \gtrsim 6$, which signals that the striped state has become unstable. The frequency of the pseudo-Goldstone mode, given by the poles in the second term of~\eqref{sigmaxxform2} are located at
\be
 \omega = - \frac{i}{2\tau_{xx}} \pm i \sqrt{\frac{1}{4\tau_{xx}^2} - \omega_{xx}^2} \ .
\ee
For positive finite $\tau_{xx}$, one of the poles lies in the upper complex $\omega$ half plane if and only if $\omega_{xx}^2<0$.   The conductivity is given by a current-current correlator, of which poles represent quasinormal modes.  If such as mode acquires a frequency with a positive imaginary part, it will be exponentially growing and lead to an instability.  However, it is not completely clear to what state this instability leads; we speculate on possibilities in Sec.~\ref{sec:summary}.

\begin{figure}[!ht]
\center
 \includegraphics[width=0.50\textwidth]{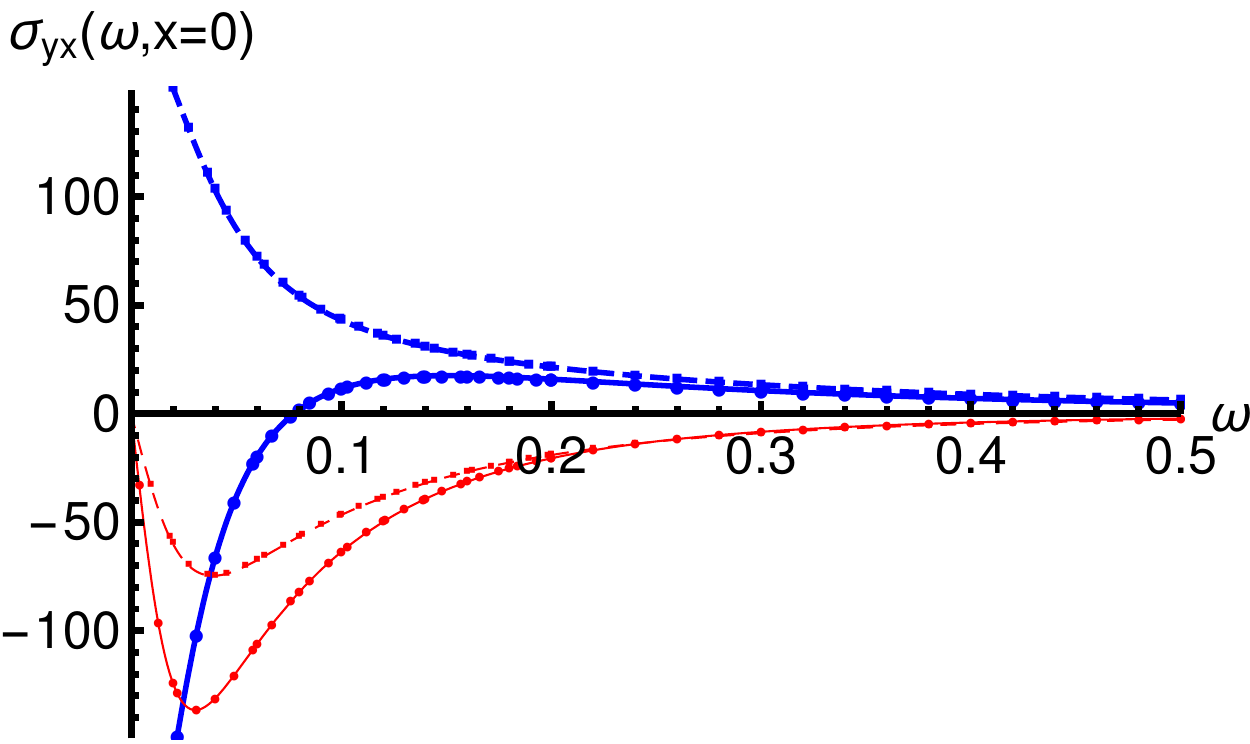}% 
 \includegraphics[width=0.50\textwidth]{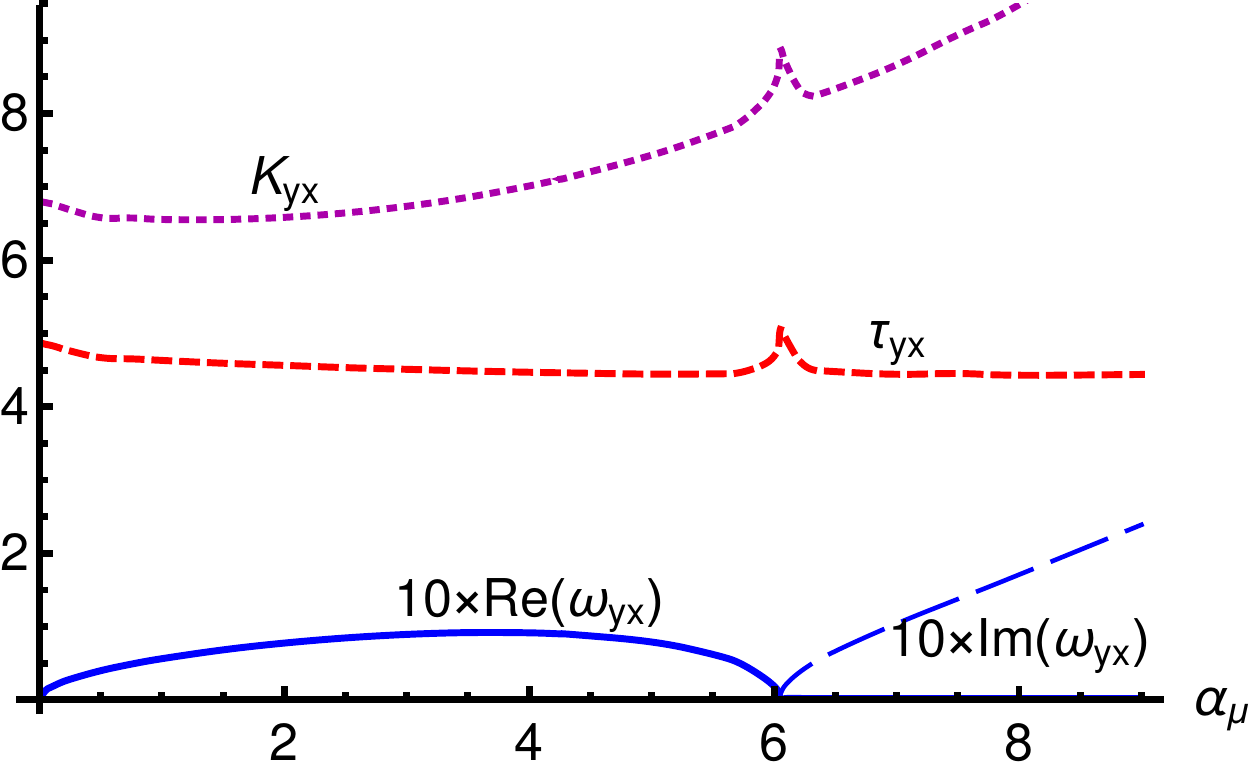}
 \caption{Fit results for the optical conductivity $\sigma_{yx}(\omega)$ at $x=0$ for ionic lattice. Left: Numerical data (dots) compared to fit results (curves). Real parts are the thick, blue curves and larger dots, and imaginary parts are the thin, red curves and smaller dots. Solid curves and round dots are at $\alpha_\mu=2$, whereas dashed curves and boxes are for $\alpha_\mu=7$. Right: The three fit parameters $K_{yx}$, $\tau_{yx}$, and $\omega_{yx}$ as functions of $\alpha_\mu$.  The resonance frequency $\omega_{yx}$ is real for $\alpha_{\mu} \lesssim 6$, and the real part is shown as a solid curve.  For $\alpha_{\mu} \gtrsim 6$, it is purely imaginary, and the imaginary part is plotted with a dashed curve. In addition, the value of $\omega_{yx}$ has been multiplied by 10 to make it visible on the same scale as the other parameters.} 
  \label{fig:sigma_yxmu_fit}
\end{figure}

Like the magnetic case, our data for the Hall conductivity $\sigma_{yx}$ can be fitted to the expression~\eqref{sigmayxform}, and the results for the fit at $x=0$ are given in Fig.~\ref{fig:sigma_yxmu_fit}. We observe that the same mode and the same instability as in $\sigma_{xx}$ also appears here: $\omega_{xx} = \omega_{yx}$, and $\tau_{xx} = \tau_{yx}$, to within the precision of the fit.\footnote{The quality of the fit is slightly worse than for $\sigma_{xx}$ because, due to the smallness of $\omega_{yx}$, the conductivity is strongly peaked near $\omega = 0$ and the peaks are not very well reproduced by our numerical data. In particular near $\alpha_\mu = 0$ and $\alpha_\mu = 6$, where $\omega_{yx}$ becomes zero, the fit contains sizable errors. This explains the bumps in $\tau_{yx}$ and $K_{yx}$ at the latter location, which are therefore identified as numerical effects.}
The overall coefficient $K_{yx}$ has a strong $x$ dependence, which is $\propto \cos(2 \pi x/L)$ at small $\alpha_\mu$, as was the case for the magnetic lattice. When $\alpha_\mu$ increases, however, higher Fourier modes set in, which is not surprising in view of the structure seen in Fig.~\ref{fig:ionic_lattice_solution}. 

\begin{figure}[!ht]
\center
 \includegraphics[width=0.50\textwidth]{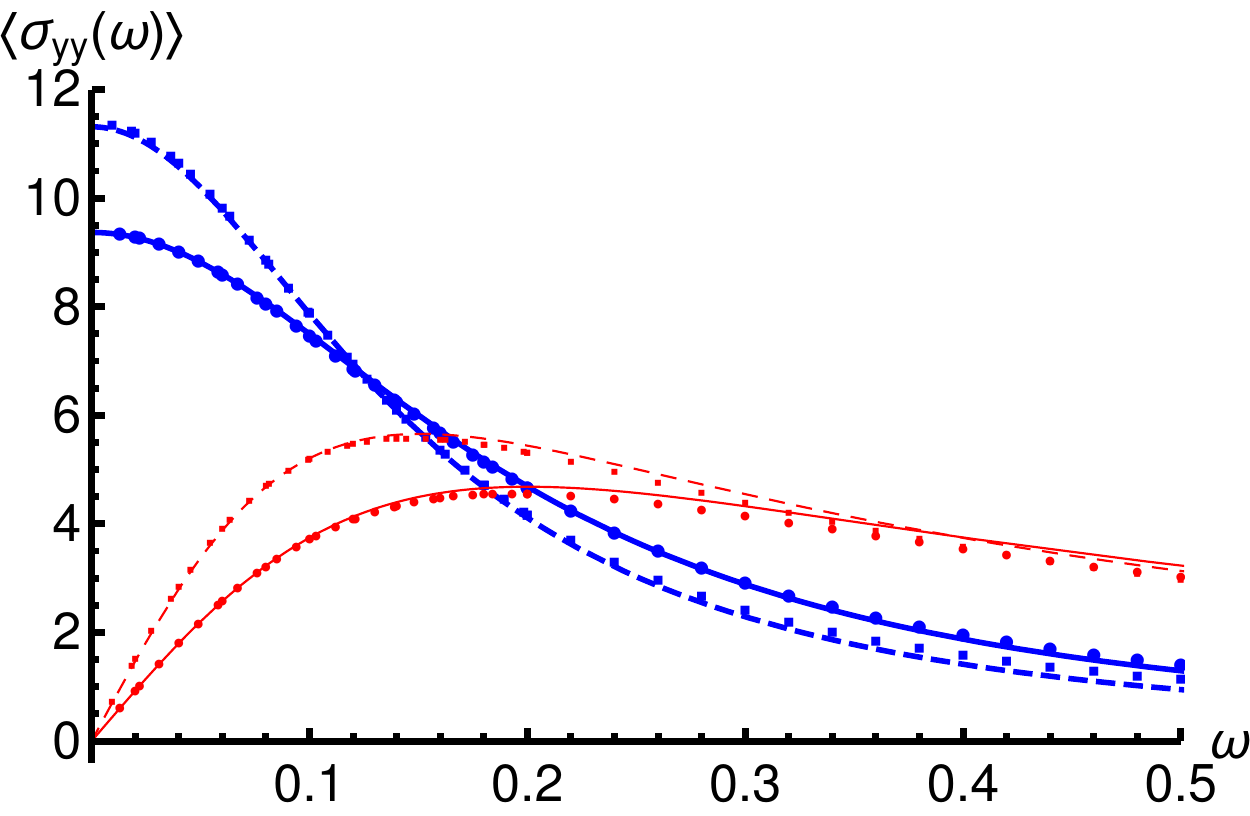}% 
 \includegraphics[width=0.50\textwidth]{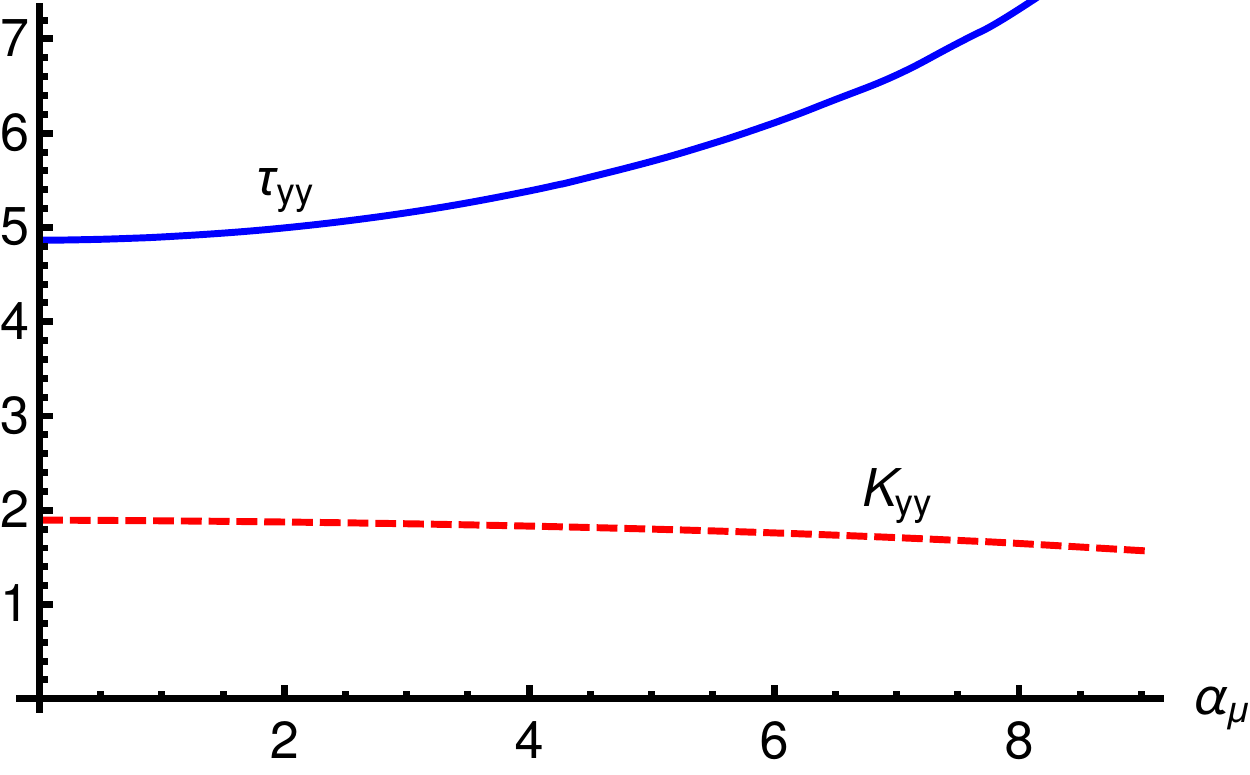}
 \caption{The averaged optical conductivity $\langle\sigma_{yy}(\omega)\rangle$. Left: The dots show the numerical data, and the curves are the results of the fit. Real parts are the thick, blue  curves and larger dots, and imaginary parts are the thin, red curves and smaller dots. Solid curves and round dots are at $\alpha_\mu=2$,  whereas dashed curves and boxes are for $\alpha_\mu=7$. Right:  The fit parameters $\tau_{yy}$ and $K_{yy}$ as functions of $\alpha_\mu$.} 
  \label{fig:sigma_yymu}
\end{figure}

As in the case of the magnetic lattice, the conductivity $\sigma_{yy}$ parallel to the stripes fits well the standard Drude form at small $\omega$ for all values of $\alpha_\mu$. 
The fit results for the ionic lattice are shown in Fig.~\ref{fig:sigma_yymu}. Essentially the only difference with respect to the results for the magnetic lattice is that the increase in $\tau_{yy}$ and the DC conductivity appears to be quadratic in the amplitude $\alpha_\mu$ of the source modulation.

%%%%%%%%%%%%%%%%%%%%%%%%%%%%%%%%%%%%%%%%%%%%%%%%%%%%%%%%%%%%%%%%%%%%%%%%%%%%%%%%%%%%%%%%%%%%%%%%%%%%%%%%%%%%%%%%%%%%%%%%%%%%%%%%%%%%%%%%%%%%%%%%%%%%%%%%%%%%%%%%%%%%%%%%%%%%%%%%%%%%%%%%%%%%%%%%%%%%%%%%%%%%%%%%%%%%%%%%%%%%%%%%%%%%%%%%%%%%%%%%%%%%%%%%%%%%%%%%%%%%%%%%%%%%%%%%%%%%%%%%%%%%%%%%%%%%%%%%%%%%%%%%%%%%%%%%%%%%%%%%%%%%%%%%%%%%%%%%%%%%%%%%%%%%%%%%%%%%%%%%%%%%%%%%%%%%%%%%%%%%%%%%%%%%%%%%%%%%%%%%%%%%%%%%%%%%%%%%%%%%%%%%%%%%%%%%%%%%%%%%%%%%%%%%%%%%%%%%%%%%%%%%%%%%%%%%%%%%%%%%%%%%%%%%%%%%%%%%%%%%%%%%%%%%%%%%%%%%%%%%%%%%%%%%%%%%%%%%%%%%%%%%%%%%%%%%%%%%%%%%%%%%%%%%%%%%%%%%%%%%%%%%%%%%%%%%%%%%%%%%%%%%%%%%%%%%%%%%%%%%%%%%%%%%%%%%%%%%%%%%%%%%%%%%%%%%%%%%%%%%%%%%%%%%%%%%%%%%%%%%%%%%%%%%%%%%%%%%%%%%%%%%%%%%%%%%%%%%%%%%%%%%%%%%%%%%%%%%%%%%%%%%%%%%%%%%%%%%%%%%%%%%%%%%%%%%%%%%%%%%%%%%%%%%%%%%%%%%%%%%%%%%%%%%%%%%%%%%%%%%%%%%%%%%%%%%%%%%%%%%

\section{Summary and future directions}\label{sec:summary} 

In previous work~\cite{Jokela:2016xuy}, we thoroughly analyzed the electric conductivities of the spontaneous striped phase. In particular, we emphasized the relevance of the sliding behavior of the stripes under an applied external electric field $E_x$ perpendicular to the modulation. 

In this paper, we introduced explicit translational symmetry breaking in the form of magnetic and ionic lattices. An immediate consequence for both types of lattices was the generation of mass for the Goldstone mode, pinning the stripes and causing an order-of-magnitude drop in the  longitudinal DC conductivity $\sigma_{xx}^{\mathrm{DC}}$ and regulating the delta peak in the Hall DC conductivity $\sigma_{yx}^{\mathrm{DC}}$. The zero-frequency limits of all optical conductivities, both for weak and strong lattices, precisely matched the analytic results derived previously in \cite{Jokela:2016xuy}.  The form of the optical conductivity $\sigma_{xx}(\omega)$ changed from a Drude peak to a Lorentzian, with a resonant peak at nonzero $\omega$, further reflecting the pinning of the stripes.

Because we only computed the linear conductivities, our calculation was not able to capture the expected  depinning transition at finite electric field. To do so, we would have to treat not only the lattice nonlinearly but the electric field as well. Our expectation is that upon constructing time-dependent solutions with {\emph{finite}} electric field, the Knonecker delta currently in the conductivity (\ref{sigmaxxDC}) will be rendered into a transition between the pinned and sliding regimes at some nonzero threshold electric field. 

To a large extent, our expectations for the magnetic lattice were met. Not only did we observe the pinning of the stripes and the suppression of $\sigma_{xx}^{\mathrm{DC}}$, but we also found that $\sigma_{yy}$ was enhanced as a function of the amplitude of the magnetic lattice $\alpha_b$.  As shown in  Fig.~\ref{fig:charge_densities}, an increase of $\alpha_b$ adds more charge to the system, and the increase in charge carriers leads to an increase in $\sigma_{yy}$.

A striking surprise was the instability associated with the ionic lattice. For weak explicit symmetry breaking, the situation was qualitatively similar to the magnetic lattice. However, for a lattice with amplitude $\alpha_\mu\sim 6$, we observed a novel instability, which was signaled by a tachyonic mode: the frequency of the pseudo-Goldstone mode entered the upper-half of the complex $\om$-plane. The fact that the unstable mode appears in the current-current correlator suggests that the charges would like to redistribute themselves. In fact, already for values of $\alpha_\mu\gtrsim 1$, there are regions of both positive and negative charges, which may be an unstable configuration; an analysis along the lines of \cite{Domokos:2013kha} might resolve this issue.

The question remains, what does this instability signal? One possibility is that the instability is due to a change in the lock-in structure: the charges would prefer to redistribute in $x$ direction so that the ground state would be modified in the IR and could be characterized by a wave number different from $k_0$. Such an instability could also be related to the appearance of higher Fourier modes in the background solution, which was seen in Fig.~\ref{fig:ionic_lattice_solution}. 
Another possibility is that the charges would prefer to redistribute in the $y$ direction, so that the translational symmetry in the $y$ direction is spontaneously broken.   In this case, the endpoint of the instability could be bubbles, i.e., coexistence of phases with different charge densities in the $y$-direction with phase boundaries, or a phase with stripes also in the $y$-direction, i.e., a checkerboard or d-wave structure. The former have been seen in  fractional quantum Hall fluids \cite{Fogler}, which very strongly resemble the system studied here.

Given the instability, it is important to extend our work to find the true ground state beyond the commensurate case studied here. We plan to investigate the phase structure with lattices of incommensurate wave number.  In addition, we aim to analyze fluctuations of the spontaneous stripes relevant for the breakdown of translational symmetry in the $y$ direction.

There are a number of other interesting directions for further research.  One involves turning on a constant external magnetic field and studying the electric transport properties of the parity-broken metallic gapless quantum fluid.  Another potential avenue would be to investigate how the striped order vanishes in the vicinity of the gapped quantum Hall state.

Generalizing the model by an $SL(2,\mathbb{Z})$ transformation \cite{Witten:2003ya,Burgess:2000kj,Burgess:2006fw,Jokela:2013hta,Jokela:2014wsa,Ihl:2016sop}\footnote{Anyonization of other D-brane models have also recently been considered \cite{Jokela:2015aha,Itsios:2015kja,Itsios:2016ffv,Jokela:2016nsv}.} allows us to address the transport of striped anyonic fluids \cite{Jokela:2017fwa}, which are notoriously difficult to tackle with perturbative methods. 

Finally, it would be interesting to try to include $1/N$ effects to model quantum phase fluctuations in order to see the transition to the bad metal phase and to connect to the recent interesting work in \cite{Delacretaz:2016ivq}.

\addcontentsline{toc}{section}{Acknowledgments}
\paragraph{Acknowledgments}

\noindent
We would like to thank  M.~Baggioli, A.~Donos, B.~ Gout\'eraux, C.~Hoyos, R.~Meyer, C.~Pantelidou, G.~Policastro, D.~Rodr\'iguez Fern\'andez, T.~Vojta,  and J.~Zaanen for discussions.  We would also like to thank T.~Andrade, M.~Baggioli,  A.~Krikun, N.~Poovuttikul, L.~Alberte, M.~Ammon, A.~Jimenez, and O.~Pujol\`as for sharing their unpublished work.
N.~J.~is supported in part by Academy of Finland Grant No. 1297472.
M.~L.~was supported by a 2017 LIU professional development grant.  M.~L.~would like to thank the KITP for hospitality and partial support via the National Science Foundation under Grant No. NSF PHY-1125915.

%%%%%%%%%%%%%%%%%%%%%%%%%%%%%%%%%%%%%%%%%%%%%%%%%%%%%%%%%%%%%%%%%%%%%%%%%%%%%%%%%%%%%%%%%%%%%%%%%%%%%%%%%%%%%%%%%%%%%%%%%%%%%%%%%%%%%%%%%%%%%%%%%%%%%%%%%%%%%%%%%%%%%%%%%%%%%%%%%%%%%%%%%%%%%%%%%%%%%%%%%%%%%%%%%%%%%%%%%%%%%%%%%%%%%%%%%%%%%%%%%%%%%%%%%%%%%%%%%%%%%%%%%%%%%%%%%%%%%%%%%%%%%%%%%%%%%%%%%%%%%%%%%%%%%%%%%%%%%%%%%%%%%%%%%%%%%%%%%%%%%%%%%%%%%%%%%%%%%%%%%%%%%%%%%%%%%%%%%%%%%%%%%%%%%%%%%%%%%%%%%%%%%%%%%%%%%%%%%%%%

\end{document}